\documentclass[aps,prd,preprint,groupedaddress]{revtex4}
\usepackage{graphicx}
\usepackage{float}                 
\usepackage{caption} 
\usepackage{array}
\usepackage{tabularx}
\usepackage{amsmath,amssymb}
\usepackage{xcolor}
\usepackage{subcaption}
\usepackage[colorlinks=true,citecolor=blue,urlcolor=blue,linkcolor=blue]{hyperref}
\usepackage{adjustbox}
\usepackage{changepage}   
\usepackage{booktabs}
\usepackage{multirow}
\usepackage[noabbrev,capitalise]{cleveref}
\crefformat{figure}{#2#1#3}
\crefrangeformat{figure}{#3#1--#2#4}
\crefmultiformat{figure}{#2#1#3}{--#2#1#3}{, #2#1#3}{, #2#1#3}

\begin{document}
	
	\title{Magnetic moments of decuplet baryons in asymmetric magnetized nuclear matter}
	\author{Utsa Dastidar$^{1}$}
	\email{utsadastidar@gmail.com}
	\author{Arvind Kumar$^{1}$}
	\email{kumara@nitj.ac.in}
	\author{Harleen Dahiya$^{1}$}
	\email{dahiyah@nitj.ac.in}
	\author{Suneel Dutt$^{1}$}
	\email{dutts@nitj.ac.in}
	\affiliation{$^1$ Department of Physics, Dr. B.R. Ambedkar National
		Institute of Technology, Jalandhar, 144008, India}

	\begin{abstract}
		Understanding the novel QCD phenomenon under high external magnetic fields of hot and dense medium help us to develop a better understanding of the underlying quark dynamics of baryons. Using a hybrid approach based on the effective field theory that treats quarks as the fundamental degrees of freedom and calculating the individual contribution of valence, sea and orbital angular moment of sea quark, the magnetic moment of a given baryon of the decuplet family is calculated. The incorporation of Landau quantization in the vector and scalar densities of baryons help us to obtain the impact of external magnetic field on the properties of baryons within the chiral SU(3) quark mean field model (CQMF). In the present study, effective masses of the baryons are calculated using CQMF while the framework of chiral constituent quark model ($\chi$CQM), extended to SU(4) sector, is used to obtain the effective magnetic moments of decuplet baryons under the influence of magnetic field.

	\end{abstract}
	
	\maketitle
	
	\section{Introduction}

	In the non-central heavy-ion collisions (HICs) \cite{Kharzeev:2007jp,Fukushima:2008xe,Skokov:2009qp,Cho:2014loa,Kharzeev:2012ph}, the produced quark-gluon plasma (QGP) evolves through various phases. The spectator ions in non-central collisions produce a magnetic field reaching magnitudes of about $eB\approx2m_\pi^2$ at the Relativistic Heavy Ion Collider (RHIC) and can attain $eB\approx15m_\pi^2$ at the Large Hadron Collider (LHC) \cite{ParticleDataGroup:2012pjm}. These conditions imitate approximately the early stages of the universe \cite{Ferris1998,Delsemme1998} which upon probing is expected to help in understanding the evolution of QGP under the induced magnetization. This strongly interacting dense hadronic matter may undergo phase transition at high temperatures/baryonic densities \cite{Pasechnik:2016wkt, Rafelski:2019twp,Heinz:2000ba}. Finite strong magnetic field may impact the nature of these phase transitions and hence, it is pivotal to understand the QCD phase diagram under the presence of an external magnetic field \cite{ATLAS:2014swk,Rafelski:2019twp}. 
	
	The static and electromagnetic properties, for example, mass and magnetic moments give us valuable information regarding the internal structure of baryons in the non-perturbative regime. Investigating the effects of finite isospin asymmetry in a strongly interacting medium under the extreme magnetic fields produced by spectator ions \cite{Haque:2019hzc} are essential, for understanding the intrinsic properties of mesons and baryons. Over the recent decades, the signatures of hot strongly interacting matter have been explored through numerous investigations. Various theoretical approaches have been employed to study these effects, particularly in relation to the electromagnetic characteristics of baryons such as magnetic moments and form factors. Notable among these are the chiral quark model \cite{Dahiya:2002qj}, quark-meson coupling model \cite{Tsushima:2019wmq}, QCD sum rules \cite{Dey:1999fi}, and light-cone QCD frameworks \cite{Aliev:2000cy}, which have been extensively used to perform such calculations in vacuum \cite{Araujo2004, HackettJones2000, Contreras2004, He:2004kg, Sahu2002, Slaughter:2011xs}. These studies contribute to a deeper understanding of hadron structure and their behavior under different QCD environments. The intense transient magnetic field generated gives us a small window to study novel QCD phenomena of breaking and restoration of chiral symmetry.
	
	The magnetic moments of octet baryons have been extensively studied theoretically \cite{Araujo2004, HackettJones2000, Contreras2004, He:2004kg} through the analyses of their internal structure \cite{Draper:1991uu}. In this context, the magnetic form factor $G_m(Q^2)$, evaluated at $Q^2 = 0$ (where $Q^2$ denotes the squared four-momentum transfer) serves as a fundamental quantity \cite{Leinweber:1990wn}. Further insights have been gained through extrapolations based on charge radii measurements \cite{Draper:1991uu} and the studies of in-medium mass modifications of baryons \cite{Ghim:2022zob}. Covariant baryon chiral perturbation theory has been applied to investigate the magnetic moments of octet baryons \cite{Geng:2008mf,Geng:2009ys}, especially incorporating the effects of SU(3) symmetry breaking. In the low-energy regime, it has been demonstrated that a chiral expansion of the magnetic moments is feasible, provided that higher-order corrections, such as small decuplet degree of freedom and loop corrections, are included \cite{Gasser1985, Gasser1988, Scherer2003, Jenkins1993, Durand1998, Puglia2000,Meissner:1997hn}. Nevertheless, to achieve a deeper and more fundamental understanding of baryonic magnetic moments, it is essential to consider the individual quark contributions underlying the composite structure \cite{EuropeanMuon:1983wih}.
	
	The MIT bag model initially offered a valuable framework for the calculation of baryonic magnetic moments by treating constituent quarks as non-interacting particles \cite{Ryu:2009yni}. Subsequent refinements introduced weak coupling effects between constituent quarks, accounting for quark interactions within baryons. Notably, the observed ratio between the contributions of the $u$- and $d$-quarks to the nucleon's magnetic moment necessitates the incorporation of dynamical quark masses for a consistent theoretical description. Consequently, the use of constituent quark masses, rather than current quark masses, becomes imperative for a realistic study of baryonic magnetic moments inclusive of quark dynamics. This necessity strongly supports the existence of relativistic and gluonic effects within baryons, phenomena which are otherwise neglected in traditional quark models.

	In order to explore the QCD phase diagram in regions that extend beyond the reach of lattice QCD regime, a variety of phenomenological models have been developed. These include the hadron resonance gas (HRG) model \cite{Huovinen:2009yb}, the Nambu–Jona–Lasinio (NJL) model, the linear sigma model (LSM) \cite{Bochkarev:1995gi}, the quark–meson (QM) model \cite{Schaefer:2004en}, and the quark–meson coupling (QMC) model \cite{Tsushima:1997cu}. Additionally, the Dyson–Schwinger approach \cite{Wadia:1980rb} and coupled–channel \cite{Waas:1996fy} frameworks grounded in fundamental QCD have also been employed to study hadronic matter under extreme conditions. To simultaneously describe the phenomena of chiral symmetry restoration and deconfinement, extensions of these models have been proposed through the incorporation of the Polyakov loop potential \cite{Fukushima:2017csk}. As a result, models such as the Polyakov–Nambu–Jona–Lasinio (PNJL) \cite{Fukushima:2003fw}, Polyakov–linear sigma model (PLSM) \cite{Mao:2009aq}, and Polyakov–quark–meson (PQM) \cite{Schaefer:2007pw} model have been established, offering a more comprehensive framework for investigating the thermodynamic \cite{Ruggieri:2016lrn,Ruggieri:2016xww} and structural properties of strongly interacting matter.
	
	In literature, the above mentioned models have explored magnetic moment of decuplet and octet baryons in free space and as well as in nuclear medium \cite{Sahu2002,Schlumpf1993, Araujo2004, HackettJones2000, Contreras2004, He:2004kg}. However, to the best of our knowledge, within the framework of the chiral constituent quark model ($\chi$CQM) the magnetic moment of the baryons under the effect of external magnetic field have not yet been studied. The absence of experimental data on magnetic moment of decuplet baryons under influence of an external magnetic field further inclines one to resort to effective models to obtain this important information. 
				
	Recent theoretical studies of strongly interacting matter under external magnetic fields have been done for the Walecka model \cite{Mukherjee:2018ebw}. Showcasing the dependence of anomalous magnetic moment in the critical temperature for phase transition under external magnetic field. The anomalous magnetic moments (AMMs) of baryons are known to have a profound influence on the behavior of strongly interacting matter in the presence of a finite external magnetic field \cite{Sinha2013,Rabhi2011,Dexheimer2012,Kumar2020a,Kumar2020b,Parui:2022msu,Mishra2020,Mishra2021,Mukherjee:2018ebw,Aguirre:2019ivr}. Their inclusion becomes particularly crucial when exploring phenomena such as magnetic catalysis (MC) and inverse magnetic catalysis (IMC), as the manifestation of these effects is closely linked to the baryonic AMMs \cite{Mukherjee:2018ebw, Aguirre:2019ivr}. Moreover, the study of AMMs is essential in astrophysical contexts, especially in understanding the internal properties of magnetars and neutron stars as these are characterized by extremely strong magnetic fields \cite{Kumari:2022hyx, Thapa:2020ohp, Marquez:2022fzh}.
	
	The chiral SU(3) quark mean field model has been used in Ref. \cite{Wang:2001hw} to examine the characteristics of isospin-symmetric strange matter, taking into account the scalar fields $\sigma$ and $\zeta$ and the vector fields $\omega$ and $\phi$. Quarks confined within baryons interact through the exchange of scalar meson fields $\sigma$, $\zeta$ and $\delta$, as well as vector meson fields $\omega$, $\rho$ and $\phi$. In earlier formulations of the model, the vector-isovector $\rho$ meson was solely considered for exploring isospin-asymmetric systems. However, subsequent studies have highlighted the importance of the scalar-isovector $\delta$ meson in capturing isospin asymmetry effects more accurately, justifying its inclusion in the present investigation \cite{Thakur:2022dxb,Li:2022okx,Liu:2001iz}. Furthermore, the presence of finite baryon density and temperature in the medium leads to modifications in the properties of constituent quarks, thereby influencing the structure and dynamics of baryons.

	In the chiral SU(4) constituent quark model the individual contributions of valence quarks, spin polarization of sea quarks and orbital angular moments contribution of sea quarks are taken into account to calculate the total magnetic moment of a given baryon including configuration mixing as well \cite{Dahiya:2023izc}. The aim of this research paper is to calculate the magnetic moment of decuplet baryons in isospin asymmetric magnetized nuclear matter. The in-medium effects and the strong interactions have been realised through the chiral SU(3) quark mean field model (CQMF) \cite{Wang:2003cna, Wang:2001hw, Wang:2001jw, Wang:2004wja} in presence of external magnetic field. Through the in-medium masses of the baryons we calculate the effective magnetic moments of the baryons using the $\chi$CQM extended to the SU(4) sector \cite{Dutt:2024lui}.
	
	This paper is organized as follows: in Sec. \ref{su3 model} the chiral SU(3) quark mean field model framework is presented. The modifications to the effective Lagrangian density and scalar and vector densities of the baryons due to presence of finite strong magnetic field are also discussed in detail in this section. The chiral constituent quark model, used to calculate the in-medium magnetic moments of decuplet baryons, is discussed in Sec. \ref{Magnetic moment XCQM}. Sec. \ref{Results} shows the detailed encapsulation of the present work with all the findings.	Finally, the paper concludes with Sec. \ref{summary}, which integrates our findings and contributions.
	
	\section{Effective masses of quarks and baryons in asymmetric nuclear matter}\label{su3 model}

	Studying the in medium properties of decuplet baryons give rise to the need of an effective field theory based on low energy properties of QCD. The chiral SU(3) quark mean field model is a suitable model for this as it considers quarks and mesons as degrees of freedom including low energy QCD properties such as the spontaneous and explicit breaking of chiral symmetry \cite{Wang:2001jw}. In this model, the quarks are confined inside the baryons by means of a confining potential. The inter hadronic interactions are described by the scalar fields $\sigma,\space\zeta$ and $\space\delta$ and vector fields $\omega$ and $\rho$ among which the $\delta$ and $\rho$ fields are to incorporate the effects of isospin asymmetry in the medium. The various interactions of the aforementioned fields are amalgamated in the Lagrangian density of the chiral SU(3) quark mean field model which is given by 
	\begin{equation} \label{eq:Leff}
		\mathcal{L}_{\text{chiral}} = \mathcal{L}_{q0} + \mathcal{L}_{qm} + \mathcal{L}_{\Sigma\Sigma} + \mathcal{L}_{VV} + \mathcal{L}_{\chi \text{SB}} + \mathcal{L}_{\Delta m} + \mathcal{L}_{c}.
	\end{equation}
	The Lagrangian for free massless quarks is given by $\mathcal{L}_{q0} = \bar{\Psi} \, i \gamma^\mu \partial_\mu \Psi,$ whereas the interaction term between quarks and mesons is described as follows \vspace{2pt}
	\begin{align}
		\mathcal{L}_{qm} &= g_s \left( \bar{\Psi}_L M \Psi_R + \bar{\Psi}_R M^\dagger \Psi_L \right) 
		- g_v \left( \bar{\Psi}_L \gamma^\mu l_\mu \Psi_L + \bar{\Psi}_R \gamma^\mu r_\mu \Psi_R \right) \nonumber \\
		&= \frac{g_s}{\sqrt{2}} \, \bar{\Psi} \left( \sum_{a=0}^{8} s_a \lambda_a + i \gamma^5 \sum_{a=0}^{8} p_a \lambda_a \right) \Psi - \frac{g_v}{2\sqrt{2}} \, \bar{\Psi} \left( \gamma^\mu \sum_{a=0}^{8} v^a_\mu \lambda_a - \gamma^\mu \gamma^5 \sum_{a=0}^{8} a^a_\mu \lambda_a \right) \Psi.\nonumber \\
	\end{align}
	The self-interactions among scalar mesons and scale-breaking effects are incorporated through the Lagrangian density, \vspace{5pt}
	\begin{equation}\label{eq:Lss}
		\mathcal{L}_{\Sigma\Sigma} = -\frac{1}{2} k_0 \chi^2 (\sigma^2 + \zeta^2 + \delta^2) 
		+ k_1 (\sigma^2 + \zeta^2 + \delta^2)^2 
		+ k_2 \left( \frac{\sigma^4}{2} + \frac{\delta^4}{2} + 3\sigma^2 \delta^2 + \zeta^4 \right)
	\end{equation}
	\begin{equation*}
		\quad + k_3 \chi (\sigma^2 - \delta^2) \zeta 
		- k_4 \chi^4 
		- \frac{1}{4} \chi^4 \ln \left( \frac{\chi^4}{\chi_0^4} \right) 
		+ \frac{\xi}{3} \chi^4 \ln \left( \left( \frac{(\sigma^2 - \delta^2) \zeta}{\sigma_0^2 \zeta_0} \right) \left( \frac{\chi^3}{\chi_0^3} \right) \right).
	\end{equation*}\\
	The trace anomaly property of QCD has been introduced in the model via the last three terms of Eq.(\ref{eq:Lss}) leading to the trace of energy momentum	tensor proportional to fourth power of dilaton field $\chi$.
	The vector meson self-interactions under mean field approximation \cite{Wang:2001hw} take the following form
	\begin{equation}
		\mathcal{L}_{VV} = \frac{1}{2} \left( \frac{\chi}{\chi_0} \right)^2 \left( m_\omega^2 \omega^2 + m_\rho^2 \rho^2 \right) 
		+ g_4 \left( \omega^4 + 6\omega^2 \rho^2 + \rho^4 \right).
	\end{equation} 
	The fifth term in Eq.(\ref{eq:Leff}), denoted as the Lagrangian density \(\mathcal{L}_{\chi SB}\), accounts for explicit chiral symmetry breaking. It takes the form
	\begin{equation}
		\mathcal{L}_{\chi SB} = \frac{\chi^2}{\chi_0^2} \left[ m_{\pi}^2 f_{\pi} \sigma + \left( \sqrt{2} m_{K}^2 f_{K} - \frac{m_{\pi}^2}{\sqrt{2}} f_{\pi} \right) \zeta \right].
	\end{equation} This term is essential in chiral effective theories as it introduces finite pseudoscalar meson masses and is in compliance with the partially conserved axial-vector current (PCAC) relations for the pion and kaon fields \cite{Wang:2001hw,Wang:2001jw,Wang:2004wja}. The scalar coupling constant \(g_s\), which is used in finding the vacuum masses of $u$ and $d$ quarks is adjusted to saturation properties of nuclear medium and as a result we obtain the masses of the $u$ and $d$ quarks to be \(m_u = m_d = 253\) MeV. To obtain a physically consistent value for the strange quark mass \(m_s\) an additional mass term \(\Delta m\) is introduced via an explicit symmetry breaking contribution (the sixth term in Eq.\eqref{eq:Leff}), as described in~\cite{Singh:2016hiw, Wang:2001jw}
	\begin{equation}
		\mathcal{L}_{\Delta m} = -(\Delta m)\, \bar{\psi} S_1 \psi.
	\end{equation} The operator \(S_1\), which selects the strange quark component is given by
	\begin{equation}
		S_1 = \frac{1}{3} \left( I - \sqrt{3} \lambda_8 \right) = \text{diag}(0, 0, 1).
	\end{equation} Using this construction, the strange quark mass in vacuum can be written as
	\begin{equation}
		m_s = -g_{\zeta}^s \zeta_0 + \Delta m,
	\end{equation}
	where \(g_{\zeta}^s\)=\(g_s\), and \(\Delta m\) is fitted to obtain \(m_s = 450\) MeV. The final term in Eq.\eqref{eq:Leff}, which effectively confines quarks within baryons, takes the following form
	\begin{equation}
		\mathcal{L}_c = -\bar{\psi} \chi_c \psi.
	\end{equation} The scalar-vector potential $\chi_c$ is denoted by 
	\begin{equation}
		\chi_c(r) = \frac{1}{4} k_c r^2 \left(1 + \gamma^0\right). 
	\end{equation} The dynamics of the quark field \(\Psi_{qi}\) inside a baryon, under the influence of a meson mean field, is governed by the Dirac equation which is expressed as
	\begin{equation}
		\left[ -i \vec{\alpha} \cdot \vec{\nabla} + \chi_c(r) + \beta m_q^* \right] \Psi_{qi} = e_q^* \Psi_{qi},
	\end{equation}
	where the indices \(q\) and \(i\) indicate the quark flavor (\(q = u, d, s\)) and the baryon type (\(i = p, n \)), respectively. The matrices \(\vec{\alpha}\) and \(\beta\) are the standard components from the Dirac formalism. 
	
	To include the impact of external magnetic field, the Lagrangian of the chiral SU(3) quark mean field model is modified by adding another term $\mathcal{L}_{mag}$ thus giving the total Lagrangian as \begin{equation}\label{eq:magL}
		\mathcal{L}_{\text{eff}} = \mathcal{L}_{\text{chiral}} + \mathcal{L}_{\text{mag}},
	\end{equation}
	where
	\begin{equation}
		\mathcal{L}_{\text{mag}} = -\bar{\psi}_i \, q_i \, \gamma_\mu \, A^\mu \, \psi_i - \frac{1}{4} \, \kappa_i \, \mu_N \, \bar{\psi}_i \, \sigma^{\mu\nu} \, F_{\mu\nu} \, \psi_i - \frac{1}{4} \, F^{\mu\nu} F_{\mu\nu}.
	\end{equation} 
	A uniform magnetic field is considered along the z-axis (plane of collision) of which the vector potential can be expressed as $\mathrm{A^\mu}=(0,0,Bx,0)$. Here, \(\psi_i\) denotes the wavefunction associated with the $i ^{th}$ baryon. The second term in the expression accounts for the tensor interaction with the electromagnetic field, represented by the field strength tensor \(F_{\mu\nu}\). This interaction term is responsible for encapsulating the effects of the anomalous magnetic moments (AMM) and includes the parameter ($\mu_N=e/(2m_N)$), which stands for the nuclear magneton. In the presence of a constant magnetic field, this contribution becomes particularly significant. The quantized nature of the energy levels for the charged particles arises as the Lorentz force comes into the picture. These discrete energy levels are called the Landau levels which come into consideration because of the interaction of both charged and uncharged baryons with the magnetic field. Thus the thermodynamic potential of the magnetized nuclear system for protons and neutrons contains separate terms for parallel and perpendicular components and is modified as \cite{Mishra:2023uhx}

	\begin{equation}\label{eq:proton_tmpot}
		\begin{split}
			\Omega^{\text{proton}}_{\text{med}} = \sum_i \Omega^i_{\text{med}} = -T \sum_i \frac{|q_i| B}{2\pi}\sum_{s=\pm1}\sum_{\nu=0}^\infty (2-\delta_{\nu0}) \int_{-\infty}^{\infty}\frac{dk_z}{2\pi} \\
			\quad\times \Bigl\{ \ln\Bigl[1+e^{-\beta(\tilde{E}^i_{\nu,s}-\mu_i^*)}\Bigr] + \ln\Bigl[1+e^{-\beta(\tilde{E}^i_{\nu,s}+\mu_i^*)}\Bigr] \Bigr\},
		\end{split}
	\end{equation}
	and
	\begin{equation}\label{eq:neutron_tmpot}
		\Omega_{\mathrm{med}}^{\mathrm{neutron}}
		= \sum_i \Omega_{\mathrm{med}}^i
		= -T \sum_i \sum_{s = \pm 1}
		\int \frac{\mathrm{d}^3k}{(2\pi)^3}
		\biggl\{
		\ln\!\bigl(1 + e^{-\beta(\tilde{E}_{s}^i - \mu_i^*)}\bigr)
		+ \ln\!\bigl(1 + e^{-\beta(\tilde{E}_{s}^i + \mu_i^*)}\bigr)
		\biggr\}.
	\end{equation} The presence of a magnetic field modifies the effective single-particle energy of charged baryons and is expressed as \cite{Mishra:2023uhx}
	\begin{equation}\label{eq:charged_baryon_energy}
		\tilde{E}_{\nu,s}^{i} = \sqrt{\left(p_{\parallel}^{i}\right)^{2} + \tilde{m}_{i}^{2}}, \quad 
	\end{equation}
	where
	\begin{equation}
		\tilde{m}_{i} = \sqrt{m_{i}^{*2} + 2\nu\,|q_{i}|B} - s\,\mu_{N}\,\kappa_{i}B.
	\end{equation} Here, the parameter \( s = \pm 1 \) denotes the spin orientation of the baryon parallel or anti-parallel with the magnetic field direction respectively and $\kappa_{i}$ denotes the anomalous magnetic moment of the baryons. The charged baryons undergo a precissional motion within the transverse plane as proposed by the Landau levels under the effect of an external magnetic field. The effective energy of uncharged baryons in presence of external magnetic field is given by 
	\begin{equation} \label{eq:neutral_baryon_energy}
		\tilde{E}_s^i = \sqrt{ \left( p_{\parallel}^i \right)^2 + \left( \sqrt{m_i^{*2} + \left( p_{\perp}^i \right)^2} - s\,\mu_N\,\kappa_i\,B \right)^2 }.
	\end{equation}
	
	The thermodynamic potential of magnetized nuclear matter is minimized with respect to the scalar fields $\sigma$, $\zeta$, and $\delta$ the dilaton field $\chi$, and the vector meson fields $\omega$ and $\rho$. 

	The system of non-linear coupled equations which give us the equations of motion in terms of the scalar and vector fields is given as \cite{Kumar:2018ujk, Kumar:2019tiw}
	\begin{align}
	k_0 \chi^2 \sigma 
	&- 4k_1 \left( \sigma^2 + \zeta^2 + \delta^2 \right) \sigma 
	- 2k_2 \left( \sigma^3 + 3\sigma \delta^2 \right) 
	- 2k_3 \chi \sigma \zeta \notag \\
	&\quad - \frac{\xi}{3} \chi^4 \left( \frac{2\sigma}{\sigma^2 - \delta^2} \right)
	+ \left( \frac{\chi}{\chi_0} \right)^2 m_\pi^2 f_\pi 
	= \sum g_{\sigma i} \rho_i^s, 
	\end{align}
	
	\begin{align}
	k_0 \chi^2 \zeta 
	&- 4k_1 \left( \sigma^2 + \zeta^2 + \delta^2 \right) \zeta 
	- 4k_2 \zeta^3 
	- k_3 \chi \left( \sigma^2 - \delta^2 \right) \notag \\
	&\quad - \frac{\xi}{3\zeta} \chi^4 
	+ \left( \frac{\chi}{\chi_0} \right)^2 
	\left[ \sqrt{2} m_K^2 f_K - \frac{1}{\sqrt{2}} m_\pi^2 f_\pi \right]
	= \sum g_{\zeta i} \rho_i^s. 
	\end{align}
	\begin{align}
		k_0 \chi^2 \delta 
		&- 4k_1 \left( \sigma^2 + \zeta^2 + \delta^2 \right) \delta 
		- 2k_2 \left( \delta^3 + 3\sigma^2 \delta \right) 
		+ 2k_3 \chi \delta \zeta \notag \\
		&\quad + \frac{2\xi}{3} \chi^4 \left( \frac{\delta}{\sigma^2 - \delta^2} \right) 
		= \sum g_{\delta i} \tau_3 \rho_i^s, 
	\end{align}
	
	\begin{equation}
		\left( \frac{\chi}{\chi_0} \right)^2 m_\omega^2 \omega 
		+ g_4 \left( 4\omega^3 + 12\rho^2 \omega \right) 
		= \sum g_{\omega i} \rho_i^v, 
	\end{equation}
	
	\begin{equation}
		\left( \frac{\chi}{\chi_0} \right)^2 m_\rho^2 \rho 
		+ g_4 \left( 4\rho^3 + 12\omega^2 \rho \right) 
		= \sum g_{\rho i} \tau_3 \rho_i^v, 
	\end{equation}
	
	\begin{align}
		k_0 \chi \left( \sigma^2 + \zeta^2 + \delta^2 \right)  
		&- k_3 \left( \sigma^2 - \delta^2 \right) \zeta 
		+ \chi^3 \left[ 1 + \ln \left( \frac{\chi^4}{\chi_0^4} \right) \right] 
		+ (4k_4 - \xi)\chi^3 \notag \\
		&\quad - \frac{4 \xi}{3} \chi^3 \ln \left( 
		\left( \frac{(\sigma^2 - \delta^2)\zeta}{\sigma_0^2 \zeta_0} \right) 
		\left( \frac{\chi}{\chi_0} \right)^3 \right) \notag \\
		&\quad + \frac{2\chi}{\chi_0^2} \left[ 
		m_\pi^2 f_\pi \sigma + \left( \sqrt{2} m_K^2 f_K 
		- \frac{1}{\sqrt{2}} m_\pi^2 f_\pi \right) \zeta \right] \notag \\
		&\quad - \frac{\chi}{\chi_0^2} \left( m_\omega^2 \omega^2 + m_\rho^2 \rho^2 \right) = 0. 
	\end{align}
			
	The vector and scalar densities of charged baryons are identified as \cite{Kumar:2018ujk,Broderick:2000pe,Broderick:2001qw}
	\begin{equation}
		\rho_i^v = \frac{|q_i|B}{2\pi^2} \left[ 
		\sum_{\nu=0}^{\nu_{\text{max}}^{(i-1)}} \int_0^{\infty} dp_{\parallel}^i \, (f_{p,\nu,s}^i - \bar{f}_{p,\nu,s}^i) 
		+ \sum_{\nu=1}^{\nu_{\text{max}}^{(i-1)}} \int_0^{\infty} dp_{\parallel}^i \, (f_{p,\nu,s}^i - \bar{f}_{p,\nu,s}^i) 
		\right],
	\end{equation}
	
	and
	
\begin{equation}
	\begin{split}
		\rho_i^s = \frac{|q_i|B m_i^*}{2\pi^2} \Bigg[ 
		& \sum_{\nu=0}^{\nu_{\text{max}}^{(i-1)}} \int_0^{\infty} \frac{dp_{\parallel}^i}{\sqrt{(p_{\parallel}^i)^2 + \tilde{m_i}^{2}}} 
		\left(1 - \frac{s \mu_N \kappa_i B}{\sqrt{m_i^{*2} + 2\nu |q_i| B}} \right) (f_{p,\nu,s}^i + \bar{f}_{p,\nu,s}^i) \\
		+ & \sum_{\nu=1}^{\nu_{\text{max}}^{(i-1)}} \int_0^{\infty} \frac{dp_{\parallel}^i}{\sqrt{(p_{\parallel}^i)^2 + \tilde{m_i}^{2}}} 
		\left(1 - \frac{s \mu_N \kappa_i B}{\sqrt{m_i^{*2} + 2\nu |q_i| B}} \right) (f_{p,\nu,s}^i + \bar{f}_{p,\nu,s}^i) 
		\Bigg].
	\end{split}
\end{equation} In the above equation, $\nu_{\text{max}}$ represents the maximum value
	of the Landau level \cite{Strickland:2012vu}. Similarly, for uncharged baryons,
	we have
	
	\begin{equation}
		\rho_i^v = \frac{1}{2\pi^2} \sum_{s=\pm1} \int_0^{\infty} p_{\perp}^i \, dp_{\perp}^i \int_0^{\infty} dp_{\parallel}^i \, (f_{p,s}^i - \bar{f}_{p,s}^i),
	\end{equation}
	
	and
	
	\begin{equation}
		\rho_i^s = \frac{1}{2\pi^2} \sum_{s=\pm1} \int_0^{\infty} p_{\perp}^i \, dp_{\perp}^i 
		\left(1 - \frac{s \mu_N \kappa_i B}{\sqrt{m_i^{*2} + (p_{\perp}^i)^2}} \right) 
		\int_0^{\infty} dp_{\parallel}^i \, \frac{m_i^*}{\tilde{E}_i^s} (f_{p,s}^i + \bar{f}_{p,s}^i).
	\end{equation} where,
	\begin{align*}
		f_{p,\nu,s}^i &= \frac{1}{1 + \exp\left[\beta\left(\tilde{E}_{\nu,s}^i - \mu_p^*\right)\right]}, \quad
		\bar{f}_{p,\nu,s}^i = \frac{1}{1 + \exp\left[\beta\left(\tilde{E}_{\nu,s}^i + \mu_p^*\right)\right]}, \\
		f_{p,s}^i &= \frac{1}{1 + \exp\left[\beta\left(\tilde{E}_{s}^i - \mu_n^*\right)\right]}, \quad
		\bar{f}_{p,s}^i = \frac{1}{1 + \exp\left[\beta\left(\tilde{E}_{s}^i + \mu_n^*\right)\right]}.
	\end{align*}
	In the above equations, the terms $f_{k,\nu,s}^{i}, \, \bar{f}_{k,\nu,s}^{i}, \, f_{k,s}^{i}, \,$ and $ \bar{f}_{k,s}^{i}$
	denote the thermal distribution functions at finite temperature for baryons and their corresponding antibaryons.
	
	The effective quark mass \(m_q^*\) is given by
	\begin{equation}\label{eq:mass}
		m_q^* = -g_\sigma^q \sigma - g_{\zeta}^q\zeta - g_\delta^q I^{3q} \delta + m_{q0},
	\end{equation}
	where \(g_q^\sigma\), \(g_q^\zeta\), and \(g_q^\delta\) are the coupling parameters for the respective mesonic fields \(\sigma\), \(\zeta\), \(\delta\) and $m_{q0}=77$ MeV for the strange quark and zero for the $u$ and $d$ quarks. The quantity \(I^{3q}\) corresponds to the 3\textsuperscript{rd} component of the isospin quantum number and is +1/2 for proton and -1/2 for neutron. The scalar and vector meson couplings are given by the following relations \cite{Kumar:2023owb}:
	\begin{align*}
		\frac{g_s}{\sqrt{2}} &= g_\delta^u = -g_\delta^d = g_\sigma^u = g_\sigma^d = \frac{1}{\sqrt{2}} g_\zeta^s, \quad
		g_\delta^s = g_\sigma^s = g_\zeta^u = g_\zeta^d = 0, \\
		\frac{g_v}{2\sqrt{2}} &= g_\rho^u = -g_\rho^d = g_\omega^u = g_\omega^d, \quad
		g_\omega^s = g_\rho^s = 0.
	\end{align*}
	
	The relation between the effective mass of the $i\space ^{th}$ baryon and the spurious center of the mass momentum $p^*_{i\space cm}$ is given by \cite{Barik:1985rm,Barik:2013lna}
	\begin{equation}\label{eq:baryonmass_chiralSU(3)}
		M_i^*=\sqrt{E_i^{*2} - <p^{*2}_{i\space cm}>}.
	\end{equation} The effective energy of the baryons $E^*_i$ is represented by the following equation 
	\begin{equation}\label{eq:baryonenergy_chiralSU(3)}
		E_i^* = \sum_q n_{qi} e_q^* + E_{i\,\text{spin}},
	\end{equation} where $n_{qi}$ represents number of \textit{q}-type quarks in $i\space ^{th}$ baryon, the $ E_{i\,\text{spin}}$ is a correction term arising from the spin-spin interaction that is fitted to obtain the masses of baryons in free space.

	\section{Magnetic moment of decuplet baryons} \label{Magnetic moment XCQM} The effective masses of quarks and baryons obtained by using CQMF model are used in the calculation of magnetic moment of decuplet baryons by using the chiral constituent quark model. The emission of an internal Goldstone Boson (GB) by the constituent quarks is the basic process in the $\chi CQM$. The emitted GB later splits into quark and anti-quark pairs.
	The interaction between quarks and the complete set of sixteen GBs comprising a 15-plet along with a singlet can be described by the effective Lagrangian
	\begin{equation}
		\mathcal{L} = g_{15} \, \bar{q} \, \Phi \, q,
	\end{equation} where \( g_{15} \) represents the coupling strength, and \( \Phi \) denotes the GB field which is given as \cite{Sharma:2010vv}

	\begingroup
	\setlength{\arraycolsep}{-1pt}
	\begin{equation}
		\Phi = \left(
		\begin{array}{cccc}
			\frac{\pi^0}{\sqrt{2}} + \beta\,\frac{\eta}{\sqrt{6}} + \zeta'\,\frac{\eta'}{4\sqrt{3}} - \gamma\,\frac{\eta_c}{4} & \pi^+ & \alpha\,K^+ & \gamma\,\overline{D}^0 \\[1ex]
			\pi^- & -\frac{\pi^0}{\sqrt{2}} + \beta\,\frac{\eta}{\sqrt{6}} + \zeta'\,\frac{\eta'}{4\sqrt{3}} - \gamma\,\frac{\eta_c}{4} & \alpha\,K^0 & \gamma\,D^- \\[1ex]
			\alpha\,K^- & \alpha\,\overline{K}^0 & -\beta\,\frac{2\eta}{\sqrt{6}} + \zeta'\,\frac{\eta'}{4\sqrt{3}} - \gamma\,\frac{\eta_c}{4} & \gamma\,D_s^- \\[1ex]
			\gamma\,D^0 & \gamma\,D^+ & \gamma\,D_s^+ & -\zeta'\,\frac{3\eta'}{4\sqrt{3}} + \gamma\,\frac{3\eta_c}{4}
		\end{array}
		\right)\nonumber .
	\end{equation}
	\endgroup
	Given the hierarchy, $m_c > m_s > m_{u,d}$ of these constraints, the SU(4) symmetry breaking is introduced along with the masses for GBs to be non-degenerate ($M_{\eta'} > M_{K,\eta} > M_{\pi}$). The parameter \( a = |g_S|^2 \) represents the transition probability associated with the chiral fluctuation process \( u(d) \to d(u) + \pi^{+(-)} \). The quantities \( a\alpha^2 \), \( a\beta^2 \), \( a\zeta^2 \), and \( a\gamma^2 \) correspond to the probabilities in decreasing order for the transitions \( u(d) \to s + K^{-(0)} \), \( u(d, s) \to u(d, s) + \eta \), and \( u(d, s) \to u(d, s) + \eta' \), respectively \cite{Sharma:2010vv, Girdhar:2015gsa}. The total magnetic moment of the baryon is calculated by taking into account the contribution of valence ($\mu_{\text{val}}^*$), sea ($\mu_{\text{sea}}^*$) and orbital angular momentum of sea quarks ($\mu_{\text{orbital}}^*$), i.e., $\mu^*_{tot} = \mu_{\text{val}}^* + \mu_{\text{sea}}^* + \mu_{\text{orbital}}^*.$  The individual contributing factor for valence and sea are given by the equation 
	\begin{equation}
		\begin{aligned}
			\mu^*_{\text{val}} &= \sum_{q = u,d,s} \Delta q_{\text{val}} \mu^*_q, \\
			\mu^*_{\text{sea}} &= \sum_{q = u,d,s} \Delta q_{\text{sea}} \mu^*_q
		\end{aligned}
	\end{equation}
	where $\Delta q_{\text{val}}$ and $\Delta q_{\text{sea}}$ represent the spin polarization of constituent valence and sea quarks respectively and their explicit expressions considering the configuration mixing can be found in Ref.\cite{Sharma:2010vv}.
	To include the impact of quark confinement along with relativistic corrections for the effective quark magnetic moment, the following expressions for the constituent quarks of a given baryon have been used \cite{Girdhar:2015gsa} 
	\begin{equation} \label{eq:indivitual_quark_magnetic_moments}
		\mu_d^* = -\left(1 - \frac{\Delta M}{M_i^*} \right), \hspace{4pt}
		\mu_s^* = -\frac{m_u^*}{m_s^*} \left(1 - \frac{\Delta M}{M_i^*} \right), \hspace{4pt}
		\mu_u^* = -2\mu_d^*, \hspace{4pt}
		\mu^*_c=-\frac{2m_u^*}{m_c}\mu_d^*.
	\end{equation} 
Here, the effective mass of baryon, $M_i^*$ is calculated using Eq.\eqref{eq:baryonmass_chiralSU(3)} and $\Delta M = M_{\text{vac}} - M_i^*, \quad \text{where } M_{\text{vac}}$ is the vacuum baryon mass. For the case of calculating the magnetic moment of orbital angular momentum of sea quarks, the following expression is used
	\begin{equation}
		\mu_{\text{orbit}} = \sum_{q=u,d,s} \Delta q_{\text{val}} \mu^{*} \left( q_{+} \rightarrow q_{-}^{\prime} \right).
	\end{equation} The orbital moment for each chiral fluctuation is represented by
	$ \mu^{*} \left( q_{+} \rightarrow q_{-}^{\prime} \right)$. These components are further scaled  by the likelihood of all such chiral variations occurring, originating from a given valence quark \cite{Dahiya:2002qj}.

	\section{Results} \label{Results}

	In this section the results on the variation in masses of the constituent quarks, decuplet baryons and effective magnetic moments of decuplet baryons in assymmetric magnetized nuclear matter under the influence of magnetic field have been discussed. The various parameters used in the present work are listed in Table 1.
%
\begin{table}[h!]
	\centering
	\renewcommand{\arraystretch}{1.4}
\begin{tabular}{|ccc|c|cc|c|ccc|cc|}
		\hline
	&	$k_0$ & & 4.94 & &$\sigma_0$ (MeV) & $-93$ & &$g_\sigma^u = g_\sigma^d$& & 2.72 &\\
		\hline
	&	$k_1$ && 2.12 & & $\zeta_0$ (MeV) & $-96.87$ && $g_\sigma^s$ && 0 &\\
		\hline
	&	$k_2$ && $-10.16$ && $\chi_0$ (MeV) & 254.4 & & $g_\zeta^u = g_\zeta^d$& & 0 & \\
		\hline
	&	$k_3$ & & $-5.38$ && $\xi$ & $\frac{6}{33}$ && $g_\zeta^s$ & & 3.84 & \\
		\hline
	&	$k_4$ & & $-0.06$ && $\rho_0$ (fm$^{-3}$) & 0.16 && $g_4$& & 37.5 & \\
		\hline
	\end{tabular}
		\caption{Model parameters and coupling constants used in the present work.}
	\end{table}
	\subsection{Quark mass behaviour in magnetized nuclear matter}\label{quark mass discussion}
	\vspace{0pt}
	\begin{figure}
		\centering
		\vspace{-6em}
		\begin{adjustbox}{center}
			\includegraphics[width=1.7\linewidth]{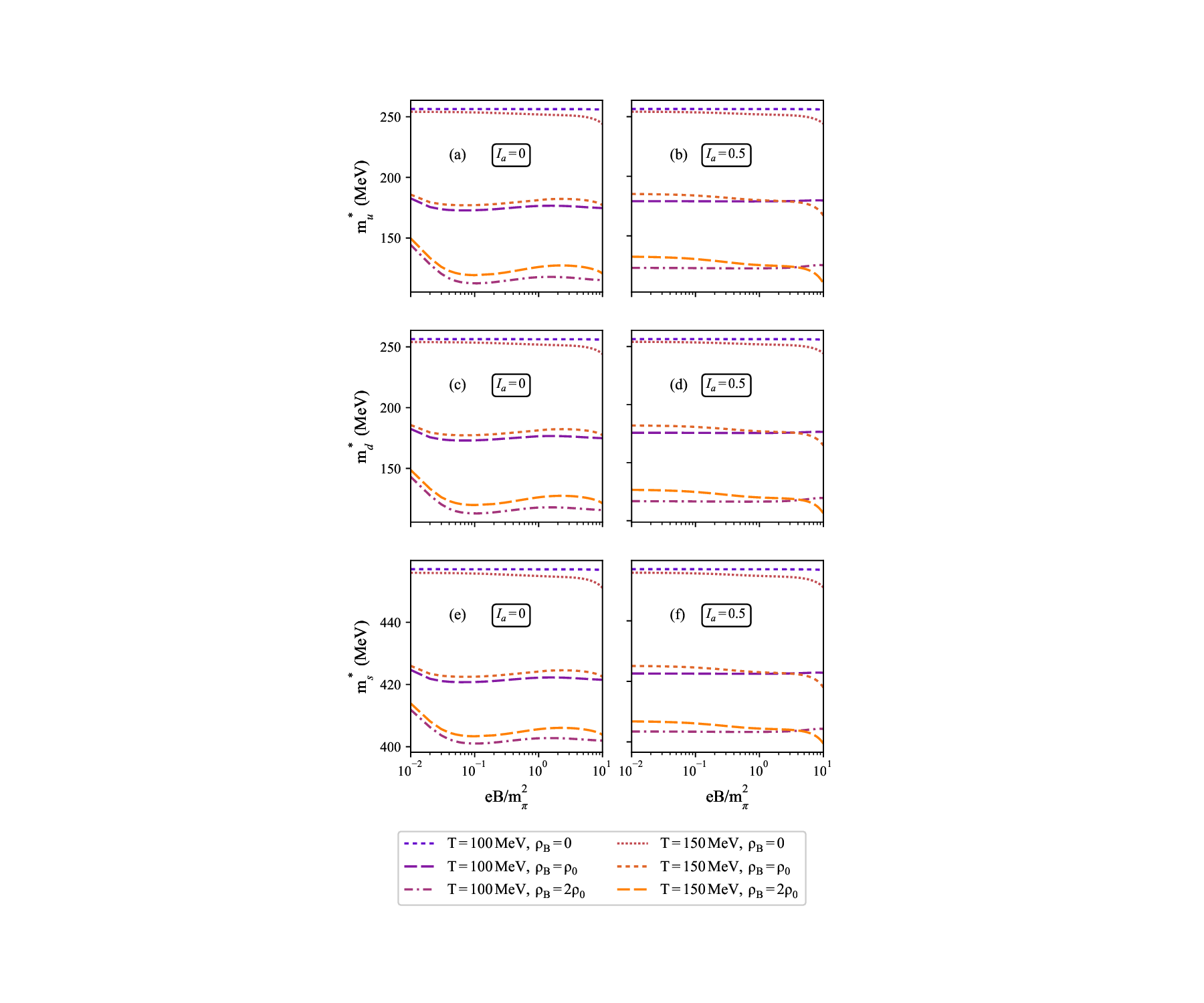}
		\end{adjustbox}\vspace{-4em}
		\caption{Effective masses of the quarks ($m_u^*, \, m_d^*$ and $m_s^*$) for isospin asymmetries $I_a=0$ [subplots (a), (c) and (e)] and $0.5$ [subplots (b), (d) and (f)], at temperatures $T=100,\,150$ MeV and baryon densities $\rho_B\,=\,0,\,\rho_0$ and $2\,\rho_0$ are shown as a function of magnetic field.}
		\label{fig:qm}
	\end{figure} In Fig. \ref{fig:qm}, the variation of effective masses $m^*_u$, $m^*_d$ and $m^*_s$ of $u$, $d$ and $s$ quarks at densities $\rho_B\,=\,0,\,\rho_0,\,2\,\rho_0$, where $\rho_0\,=\,0.16\,\mathrm{fm^{-3}}$, temperatures T = 100 and T = 150 MeV at $I_a=0$ [in subplots (a), (c) and (e)] and $I_a\,=\,0.5$ [in subplots (b), (d) and (f)], are shown as a function of external magnetic field.
	
	At zero magnetic field, baryon density $\rho_B\,=\,0$, $I_a\,=\,0$ and temperature T = 100 MeV, the mass of $u$ quark is calculated to be 256 MeV which is same as of $d$ quark as well, the mass of $s$ quark is found to be 457 MeV. A decrease in the mass of quarks is observed as density is increased from $\rho_B\,=\,0$ to $\rho_0$, showing a drop of 29\% for the light quarks and 7\% for the $s$ quark. Further increasing the density to 2$\rho_0$, we observe change of 43\% and 10\% for light and heavy quarks when compared to their values at zero baryonic density respectively. 
	As a function of magnetic field, the effective mass of quarks at $\rho_B\,=\,0$ for T = 100 MeV does not change with increasing magnetic field. An increase in temperature to T = 150 MeV shows negligible variance at higher values of magnetic field. For finite values of density, i.e., at $\rho_B\,=\,\rho_0$ and $2\rho_0$ and temperature T = 150 MeV, we observe no significant variation occurring for the change in temperature at a particular density. 
	At finite density, $\rho_B\,=\,\rho_0$, temperatures T = 100, 150 MeV and $I_a\,=\,0$ a decrease is observed in Figs. \cref{fig:qm}(a), (c), (e) at around $eB\,=\,0.07\,m_\pi^2$ showing a dip of about $5\%$ for both $u$ and $d$ quark and around $1\%$ for $s$ quark when compared with their values of mass in the absence of an external magnetic field. Further increasing the density from $\rho_0$ to $2\,\rho_0$ a higher change is observed, thus showing a decrease $\approx$ $19\%$ for $m_u,\,m_d$ and $2\%$ for $m_s$. This decrease is reflected in the masses and the magnetic moment of the decuplet baryons as well. This dip tends to disappear at lower densities and finite value of isospin asymmetry factor ($I_a$). 
	
	
	For the case of finite isospin asymmetry factor, $I_a=0.5$ in Figs. \cref{fig:qm}(b), (d), (f) at any particular temperature and in the absence of magnetic field, the change in mass shows similar variance to that of symmetric matter with increasing densities to $2\,\rho_0$. The $u$ and $d$ quark show a decrease of 30\% and 32\% in mass with increase in density from $\rho_B\,=\,0$ to $\rho_0$ and a further drop of 31\% and 32\% is shown when density is increased to $2\,\rho_0$, respectively. This observation shows a slight difference in decrease for the u quark and the d quark which is unlike the results of symmetric matter. The scalar-isovector $\delta$ meson field has a non-zero value for interaction with asymmetric matter. The coupling of $\delta$ with light quark doublet is the most significant reason behind the mass difference of $u$ and $d$ quarks. 
 	With increasing magnetic field strengths to $10 \, m_\pi^2$ no significant change is observed for T = 100 MeV at zero baryonic density. Unlike symmetric matter, no such dip is observed at $0.07 \, m_\pi^2$ for asymmetric matter at any calculated density. A relative drop of $11\,\%$ and $17\%$ is observed for $u$ and $d$ quark masses at 10 $m_\pi^2$ when temperature is increased to T = 150 MeV from 100 MeV for densities $\rho_B \,=\, \rho_0$ and $2\, \rho_0$ at $I_a\,=\,0.5$ when compared with their values at zero magnetic field, respectively. However, for the $s$ quark a lower relative change is observed, when compared to the light quarks, of about $1\%$ for the finite densities which is due to the scalar-isovector field $\delta$ having zero coupling with $s$ quark. 
	\subsection{Effective masses of decuplet baryons in magnetized asymmetric nuclear matter}\label{mass discussion}
\begin{figure}[H]
	\centering
	\includegraphics[width=.7\linewidth, trim=0cm 2.5cm 0cm 0cm, clip]{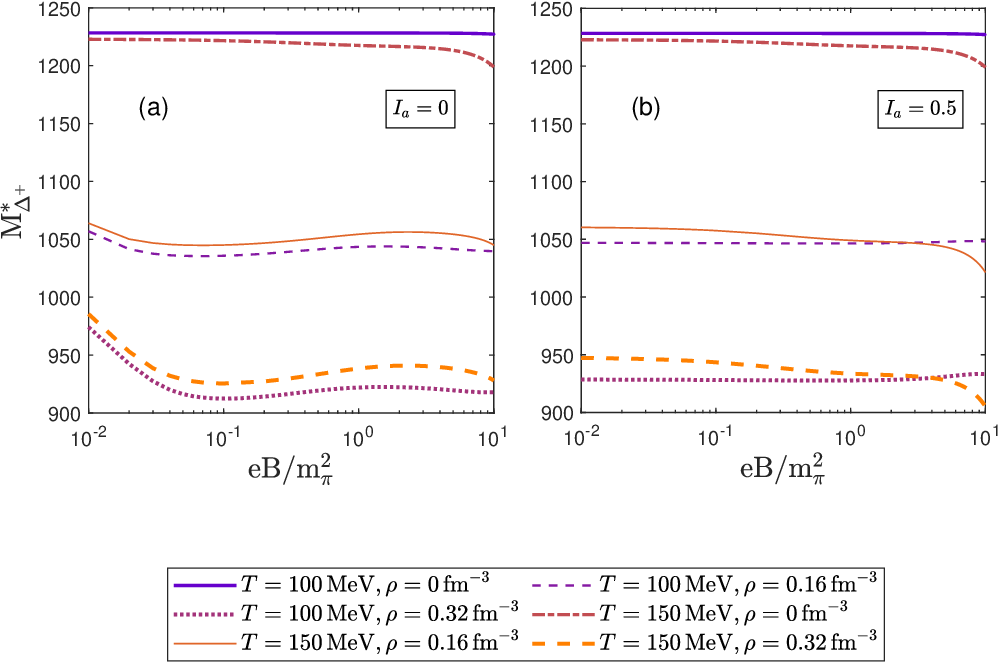}
	
	\includegraphics[width=.7\linewidth, trim=0cm 0cm 0cm 0cm, clip]{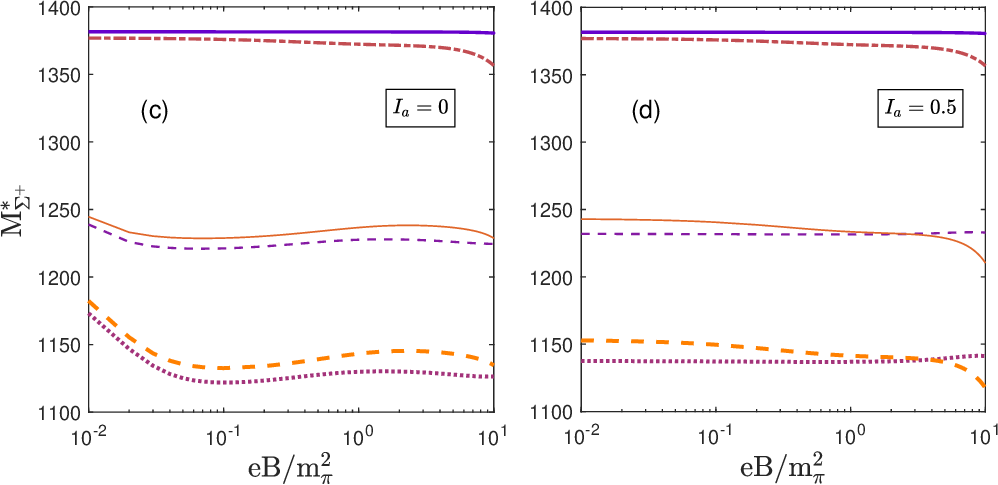}
	
	\includegraphics[width=0.7\linewidth, trim=0cm 2.5cm 0cm 0cm, clip]{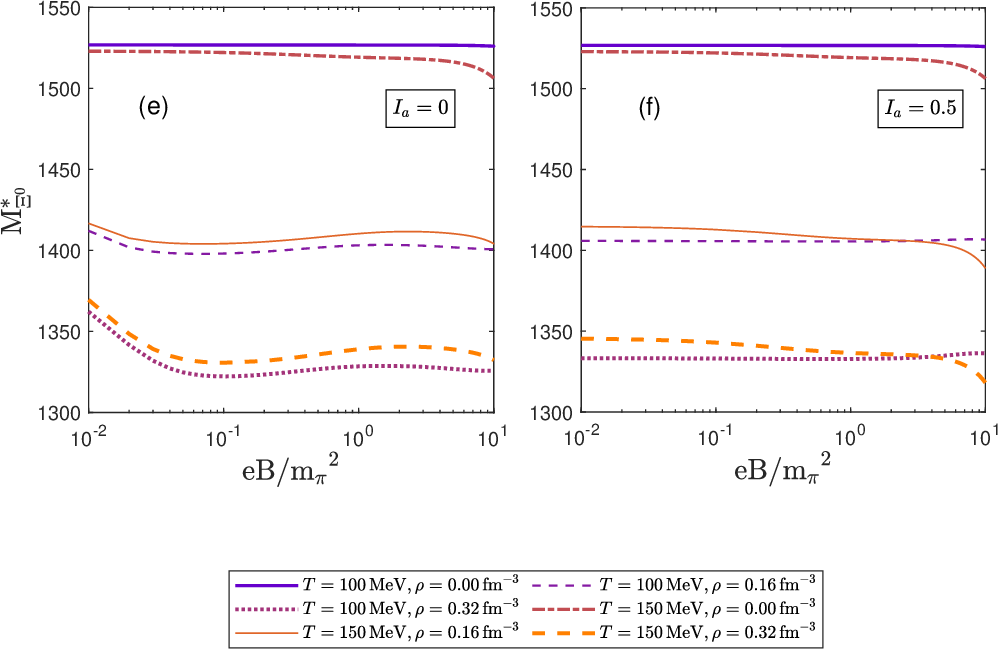}\vspace{2mm}
	\hspace*{1cm}\includegraphics[width=0.55\linewidth,trim=0.0cm 0.0cm 0.0cm 0.0cm,clip]{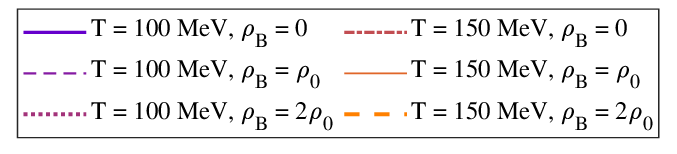}
	\caption{Mass variation of $\Delta^+$, $\Sigma^{*+}$ and $\Xi^{*0}$ is shown for isospin asymmetries $I_a\,=\,0$ [subplots (a),(c) and (e)] and 0.5 [subplots (b),(d) and (f)] at temperatures T = 100 and 150 MeV and baryon densities $\rho_B\,=\,0,\,\rho_0$ and $2\,\rho_0$ as a function of $eB/m_\pi^2$.}\label{fig:Xi_mass_log}
\end{figure}

	The effective masses of quarks are used to calculate the masses of the decuplet baryons using the Eq. \eqref{eq:baryonmass_chiralSU(3)}. The effective masses of decuplet baryons have been shown as a function of the magnetic field ($eB/m_\pi^2$) for temperatures T = 100, 150 MeV and isospin asymmetry factor $I_a=0$ (left panel)$,\,0.5$ (right panel). In each of the plots shown in Fig. \cref{fig:Xi_mass_log}, the results are shown for densities $\rho_B=0,\,\rho_0,\,2\,\rho_0$. The results found for the variation in mass of quarks, as realized through the Eq. \eqref{eq:mass}, transcends to the behavior of baryon masses with respect to magnetic field as well. In symmetric matter, i.e., $I_a\,=\,0$, Figs. \cref{fig:Xi_mass_log} (a), (c), (e), at T = 100 MeV and zero magnetic field, the effective masses of baryons decrease with increasing baryonic density similar to that observed for quark masses. A decrease of $14\%,\,10\%$ and $8\%$ is observed for the $\Delta^+,\,\Sigma^{*+}$ and $\Xi^{*0}$ respectively as density is increased from $\rho_B\,=\,0$ to $\rho_0$. An increase in density to $2\,\rho_0$ results in further decrease of $8\%,\,5\%$ and $4\%$ for the baryons $\Delta^+,\,\Sigma^{*+}$ and $\Xi^{*0}$, respectively.
	As magnetic field is increased to $10\,m_\pi^2$, the effective mass of baryons are observed to be slightly decreasing in nature when compared to values at zero magnetic field. For T = 100 MeV in symmetric matter, at $\rho_B\,=\,0$ and $\rho_0$, the trend shows almost no significant change with respect to magnetic field, at density $2\,\rho_0$ a slight decrease of about $6\%,\,4\%$ and $3\%$ is observed at $10\,m_\pi^2$ when compared with values at zero magnetic field for the baryons $\Delta^+,\,\Sigma^{*+}$ and $\Xi^{*0}$, respectively. Hence, baryons with lighter constituent quarks reflect more decrease with respect to the increase in magnetic field while heavier baryons show little to no variance. It is observed in Fig. \cref{fig:Xi_mass_log} [subplots (a), (c) and (e)], that with increasing magnetic field the mass variation shows most of the decline at lower magnetic field ($\approx 0.07\,m_\pi^2$) for finite densities after which increasing the magnetic field strength the masses show almost negligible variance. For the case of $\rho_B\,=\,2\,\rho_0$ in Fig. \cref{fig:Xi_mass_log}(a), a dip is found at $eB\,\approx\,0.07\,\,m_\pi^2$ and this nature is seen to be consistent for all baryons. The magnitude of this dip in effective masses when compared with values at zero magnetic field, reveals that it is dependent on the type of constituent quarks showing a dip of  $7\%,\,4\%$ and $3\%$ for $\Delta^+,\,\Sigma^{*+}$ and $\Xi^{*0}$, respectively. Lighter the constituent quarks, more the observed dip was. Thus at such lower strengths of magnetic field, the effect of finite densities depict a sensitive nature of the effective masses of the baryons. Increasing the temperature to T = 150 MeV also results in a slight drop in mass compared to that of T = 100 MeV at any specific magnetic field at $\rho_B\,=\,0$. However, the effective mass observed for finite density and T = 150 MeV corresponds for a slightly lower value compared to T = 100 MeV for any particular magnetic field strength.
	
	For finite isospin asymmetry factor $I_a=0.5$ presented in Figs. \cref{fig:Xi_mass_log}, subfigures labelled (b),(d) and (f), in the absence of magnetic field, the effective mass of the baryons decrease with the increase in baryonic density which is similar to the results obtained for $I_a\,=\,0$ and is consistent for all the decuplet baryons. The decrement is marginally increased as compared to $I_a=0$, showing $11\%,\,7\%$ and $5\%$ for $\Delta^+,\,\Sigma^{*+}$ and $\Xi^{*0}$, respectively. The increase in magnetic field from zero to $10\,m_\pi^2$ shows barely any change for the trends of effective mass of baryons in asymmetric matter at T = 100 MeV for all densities. However, for T = 150 MeV a slight decrease in mass is shown at higher magnetic field strength at a fixed density. For $I_a\,=\,0.5$, no significant change is observed till $\approx 2.5\,m_\pi^2$ and $4\,m_\pi^2$ for $\rho_B\,=\,\rho_0$ and $2\,\rho_0$ respectively, after which the trend of T = 100 MeV shows remains same while trend for T = 150 MeV shows a decrease till $10\,m_\pi^2$ for a particular finite density, unlike the case of $I_a\,=\,0$. A mild increase in the mass till $10\,m_\pi^2$ for T = 100 MeV and $\rho_B\,=\,\rho_0$ and $2\,\rho_0$ is observed, showing an increase of $\approx$ 2 MeV in mass. This increasing region is seen in the mass variation for all the decuplet baryons. 
		
	\subsection{Magnetic moment of decuplet baryons in magnetized asymmetric nuclear matter} \label{magnetic moment discussion}
\begin{figure}[H]
	\centering
	\includegraphics[width=1\textwidth]{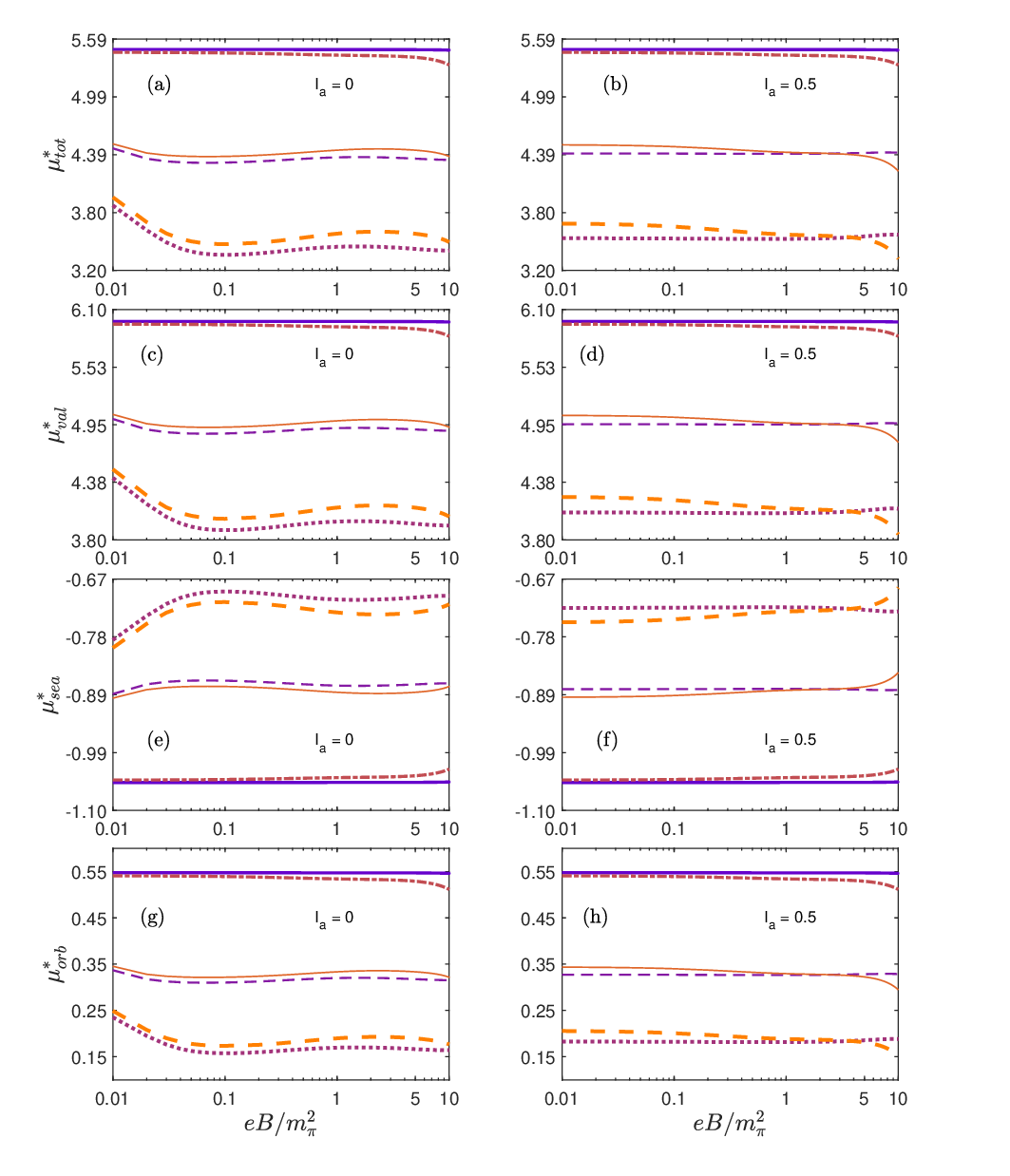}\\[-2ex]
	\includegraphics[width=0.55\textwidth]{Fig_legends.eps} \\[-3ex]
	\caption{The individual magnetic moment contributions from valence, sea and orbital moment of sea quarks to the total magnetic moment of \(\Delta^{++}\) for \(I_a = 0\) (left hand panel) and \(I_a = 0.5\) (right hand panel) shown as a function of magnetic field from 0 to 10 \(eB/m_\pi^2\) at baryonic densities $\rho_B \,=\,0,\,\rho_0$ and $2\,\rho_0$.}
	\label{fig:DeltaPP_magmom}
\end{figure}

\begin{figure}[H]
	\centering
	\includegraphics[width=1\textwidth]{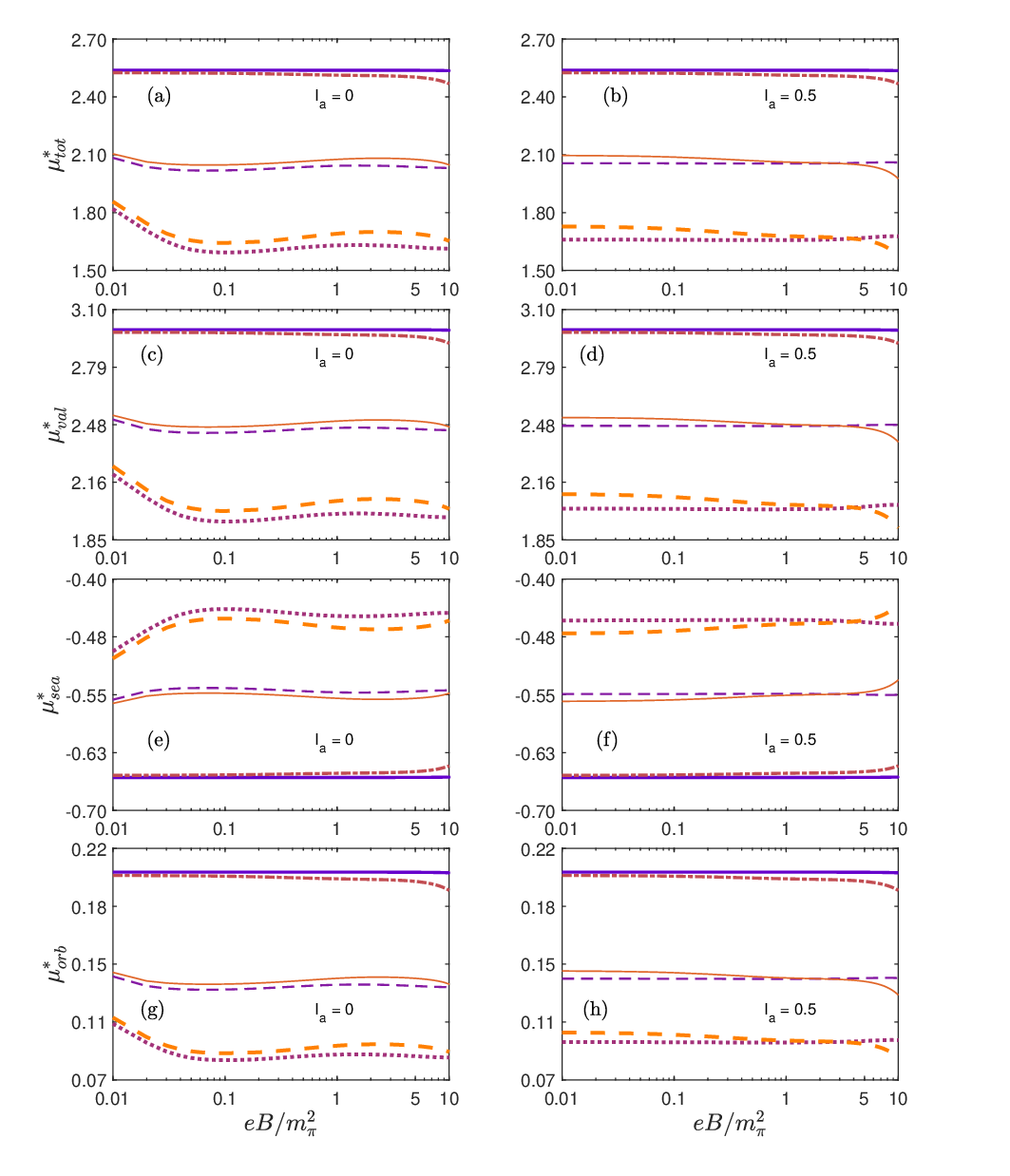}\\[-2ex]
	\includegraphics[width=0.55\textwidth]{Fig_legends.eps} \\[-3ex]
	\caption{The individual magnetic moment contributions from valence, sea and orbital moment of sea quarks to the total magnetic moment of \(\Delta^+\) for \(I_a = 0\) (left hand panel) and \(I_a = 0.5\) (right hand panel) shown as a function of magnetic field from 0 to 10 \(eB/m_\pi^2\) and baryonic densities $\rho_B \,=\,0,\,\rho_0$ and $2\,\rho_0$.}
	\label{fig:DeltaP_magmom}
\end{figure}

\begin{figure}[H]
	\centering
	\includegraphics[width=1\textwidth]{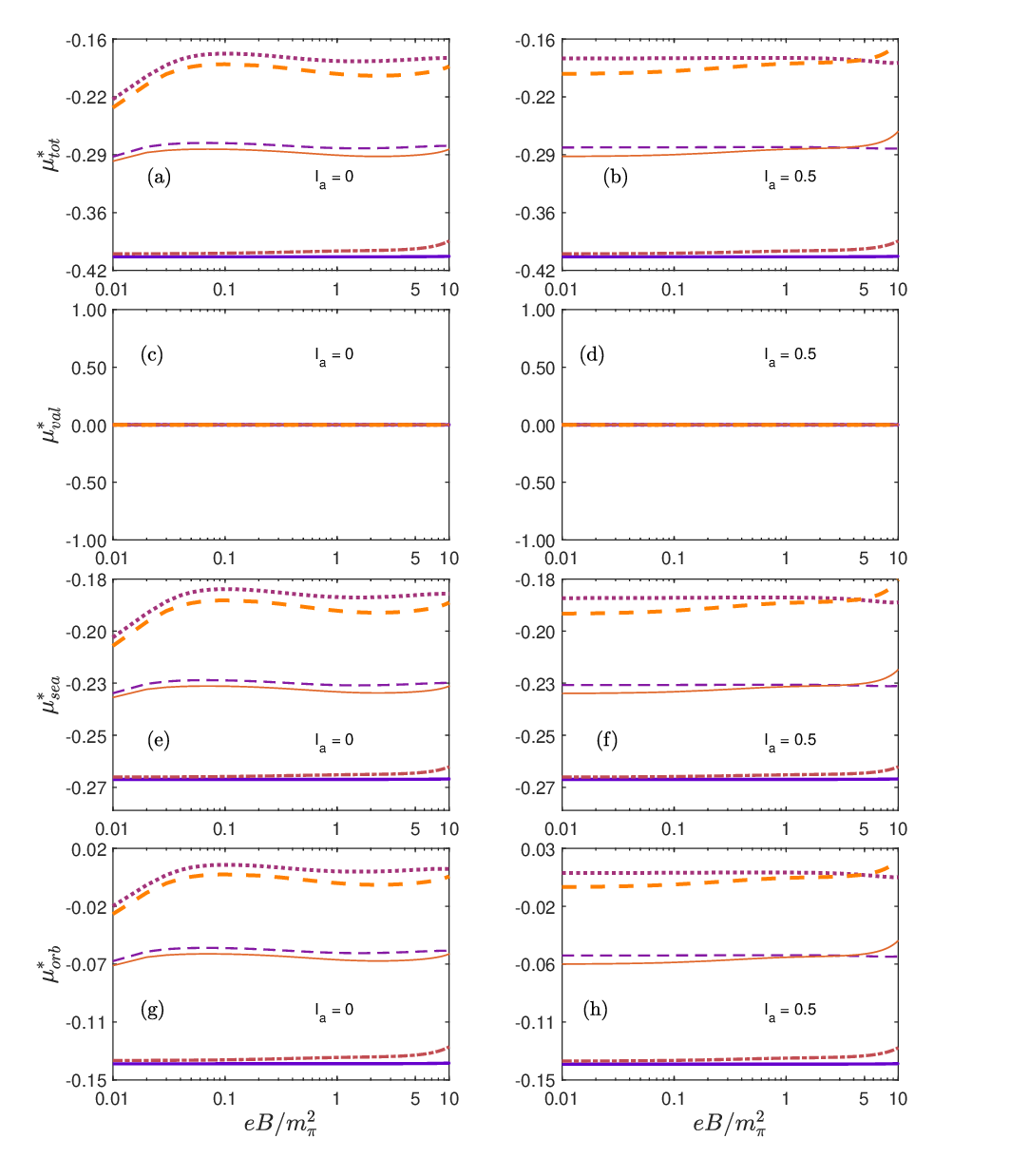}\\[-2ex]
	\includegraphics[width=0.55\textwidth]{Fig_legends.eps} \\[-3ex]
	\caption{The individual magnetic moment contributions from valence, sea and orbital moment of sea quarks to the total magnetic moment of \(\Delta^0\) for \(I_a = 0\) (left hand panel) and \(I_a = 0.5\) (right hand panel) shown as a function of magnetic field from 0 to 10 \(eB/m_\pi^2\) and baryonic densities $\rho_B \,=\,0,\,\rho_0$ and $2\,\rho_0$.}
	\label{fig:Delta0_magmom}
\end{figure}

\begin{figure}[H]
	\centering
	\includegraphics[width=1\textwidth]{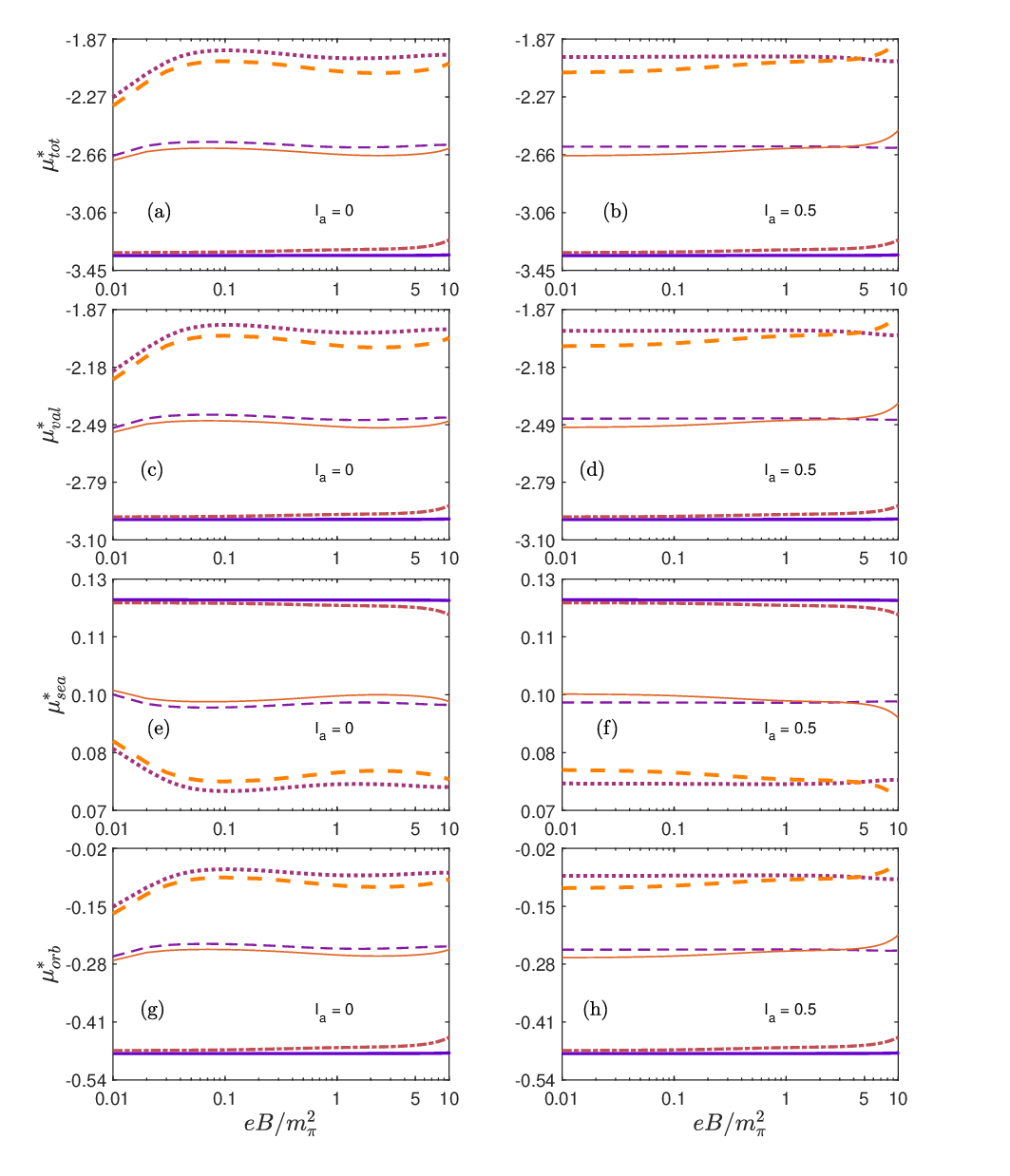}\\[-2ex]
	\includegraphics[width=0.55\textwidth]{Fig_legends.eps} \\[-3ex]
	\caption{The individual magnetic moment contributions from valence, sea and orbital moment of sea quarks to the total magnetic moment of \(\Delta^-\) for \(I_a = 0\) (left hand panel) and \(I_a = 0.5\) (right hand panel) shown as a function of magnetic field from 0 to 10 \(eB/m_\pi^2\) and baryonic densities $\rho_B \,=\,0,\,\rho_0$ and $2\,\rho_0$.}
	\label{fig:DeltaM_magmom}
\end{figure}

\begin{figure}[H]
\centering
\includegraphics[width=1\textwidth]{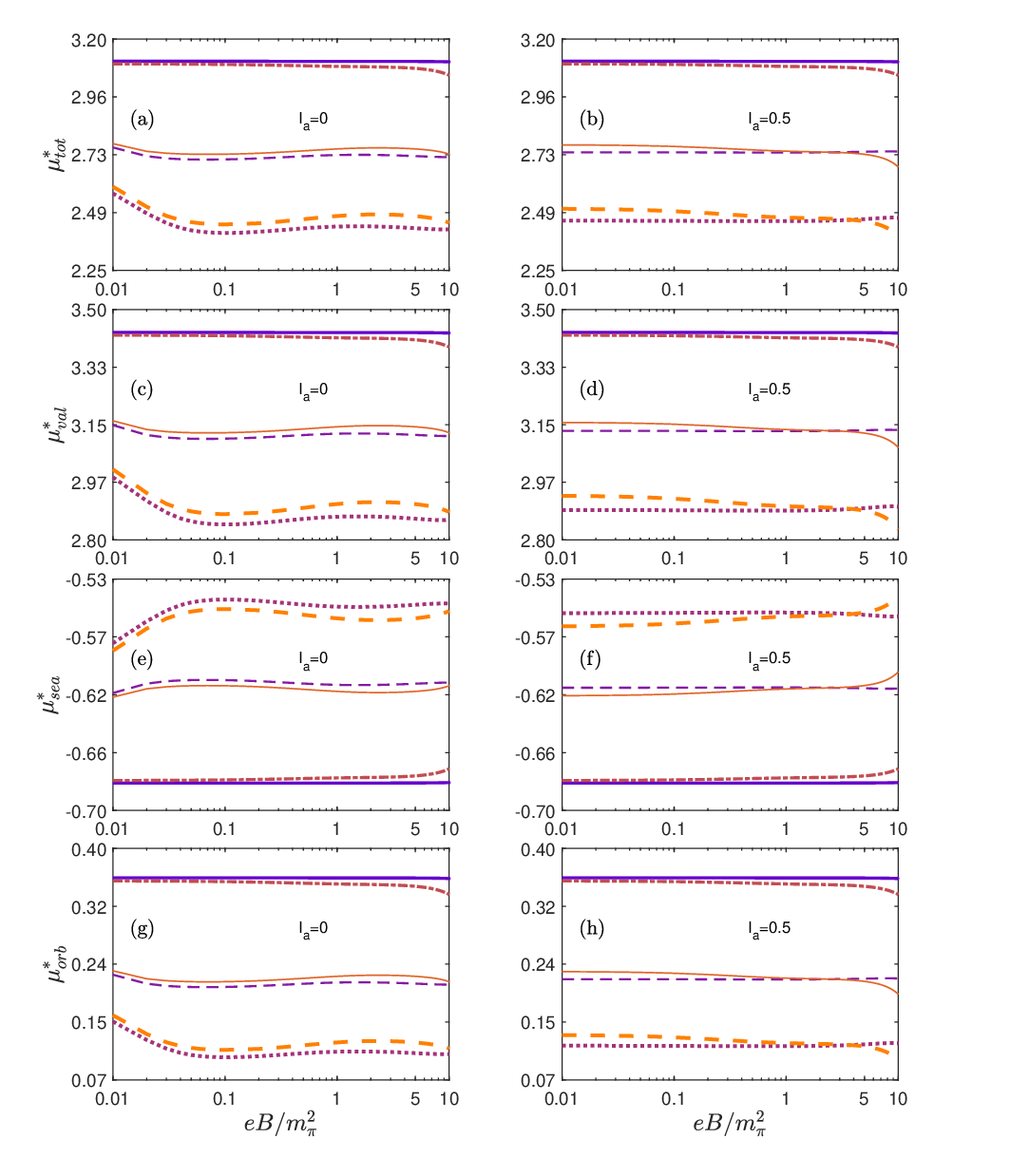}\\[-2ex]
\includegraphics[width=0.55\textwidth]{Fig_legends.eps} \\[-3ex]
\caption{The individual magnetic moment contributions from valence, sea and orbital moment of sea quarks to the total magnetic moment of \(\Sigma^{*+}\) for \(I_a = 0\) (left hand panel) and \(I_a = 0.5\) (right hand panel) shown as a function of magnetic field from 0 to 10 \(eB/m_\pi^2\) and baryonic densities $\rho_B \,=\,0,\,\rho_0$ and $2\,\rho_0$.}
\label{fig:SigmaP_magmom}
\end{figure}

\begin{figure}[H]
\centering
\includegraphics[width=1\textwidth]{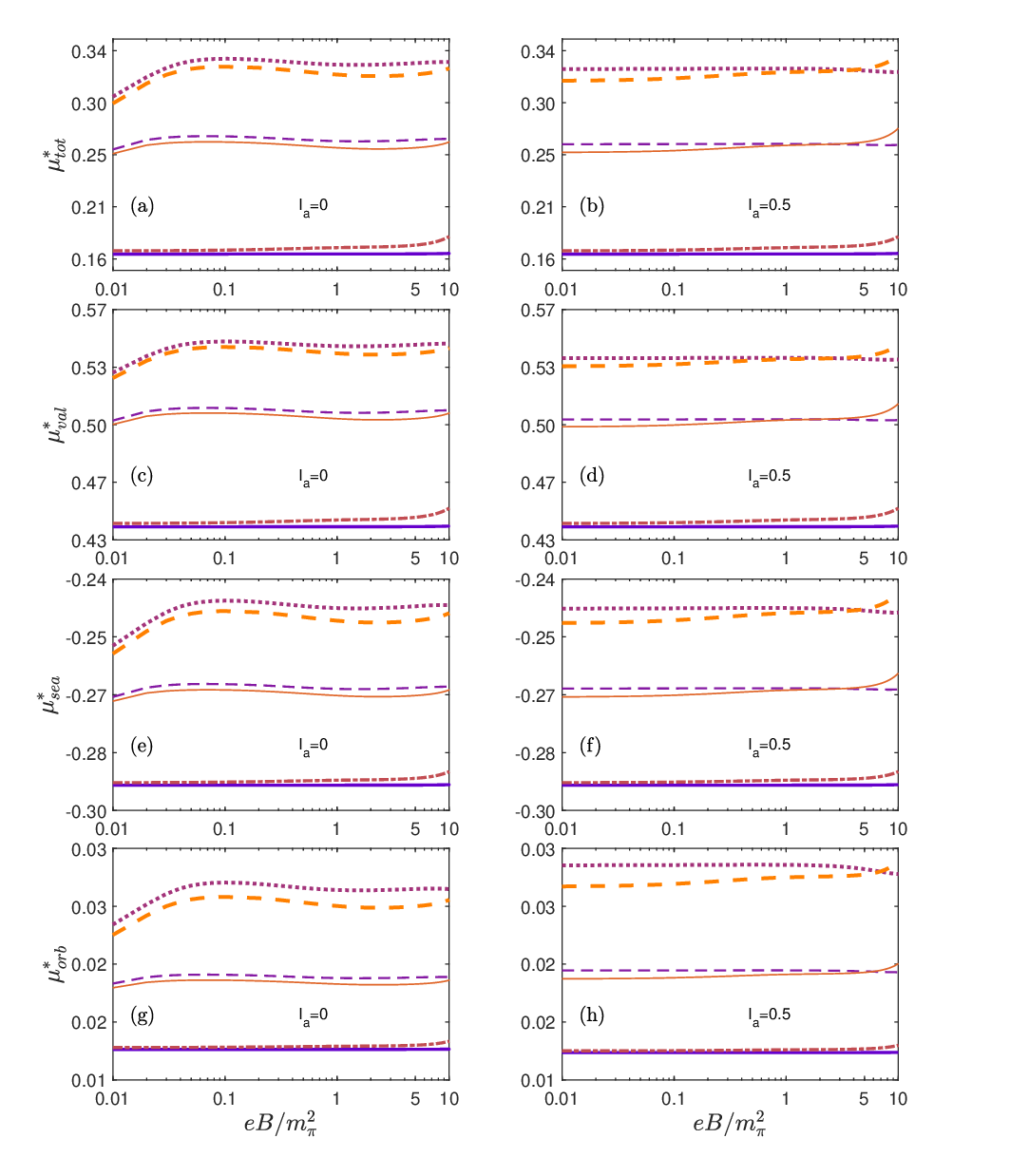}\\[-2ex]
\includegraphics[width=0.55\textwidth]{Fig_legends.eps} \\[-3ex]
\caption{The individual magnetic moment contributions from valence, sea and orbital moment of sea quarks to the total magnetic moment of \(\Sigma^{*0}\) for \(I_a = 0\) (left hand panel) and \(I_a = 0.5\) (right hand panel) shown as a function of magnetic field from 0 to 10 \(eB/m_\pi^2\) and baryonic densities $\rho_B \,=\,0,\,\rho_0$ and $2\,\rho_0$.}
\label{fig:Sigma0_magmom}
\end{figure}

\begin{figure}[H]
\centering
\includegraphics[width=1\textwidth]{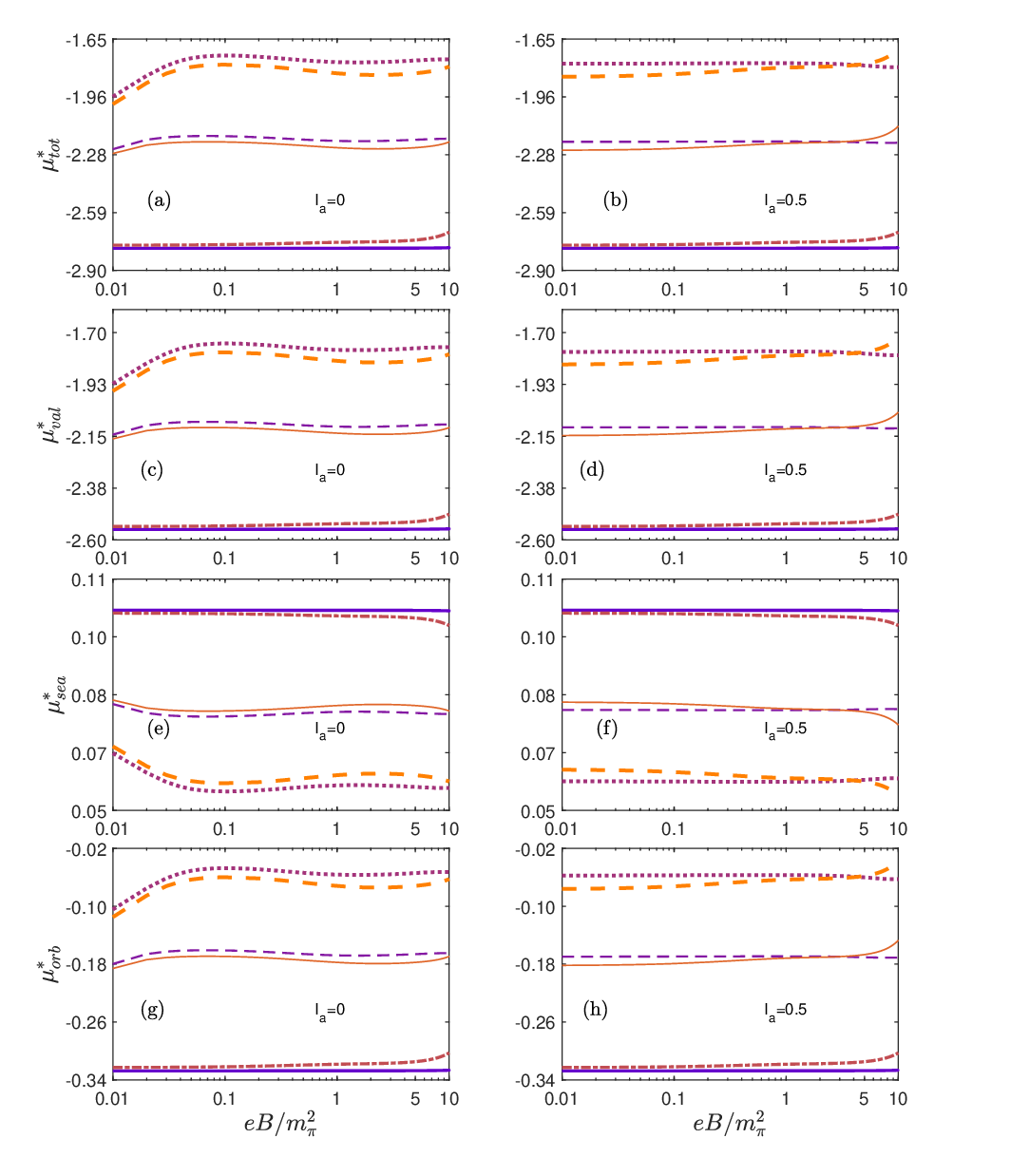}\\[-2ex]
\includegraphics[width=0.55\textwidth]{Fig_legends.eps} \\[-3ex]
\caption{The individual magnetic moment contributions from valence, sea and orbital moment of sea quarks to the total magnetic moment of \(\Sigma^{*-}\) for \(I_a = 0\) (left hand panel) and \(I_a = 0.5\) (right hand panel) shown as a function of magnetic field from 0 to 10 \(eB/m_\pi^2\) and baryonic densities $\rho_B \,=\,0,\,\rho_0$ and $2\,\rho_0$.}
\label{fig:SigmaM_magmom}
\end{figure}

\begin{figure}[H]
	\centering
	\includegraphics[width=1\textwidth]{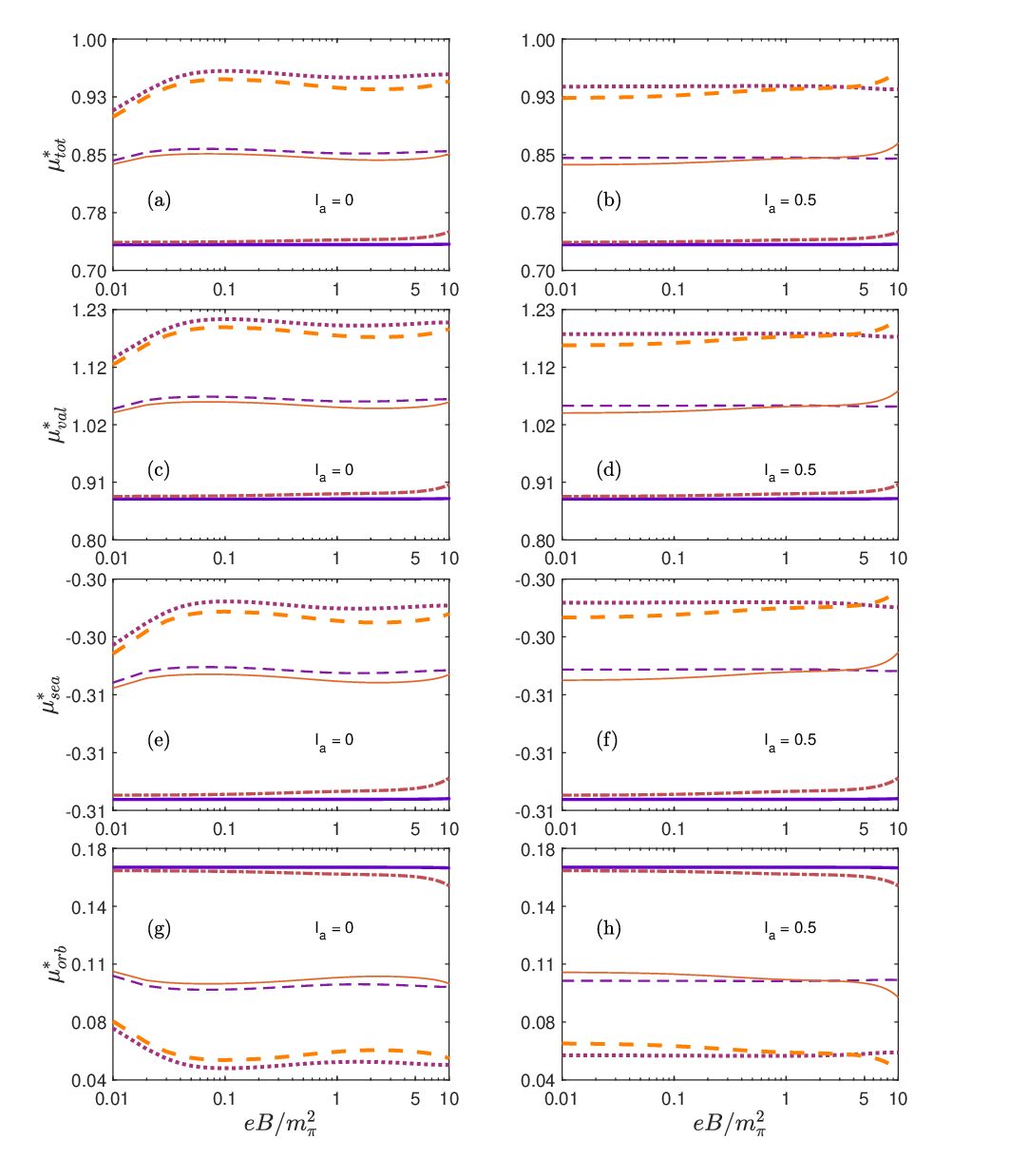}\\[-2ex]
	\includegraphics[width=0.55\textwidth]{Fig_legends.eps} \\[-3ex]
	\caption{The individual magnetic moment contributions from valence, sea and orbital moment of sea quarks to the total magnetic moment of \(\Xi^{*0}\) for \(I_a = 0\) (left hand panel) and \(I_a = 0.5\) (right hand panel) shown as a function of magnetic field from 0 to 10 \(eB/m_\pi^2\) and baryonic densities $\rho_B \,=\,0,\,\rho_0$ and $2\,\rho_0$.}
	\label{fig:Xi0_magmom}
\end{figure}

\begin{figure}[H]
	\centering
	\includegraphics[width=1\textwidth]{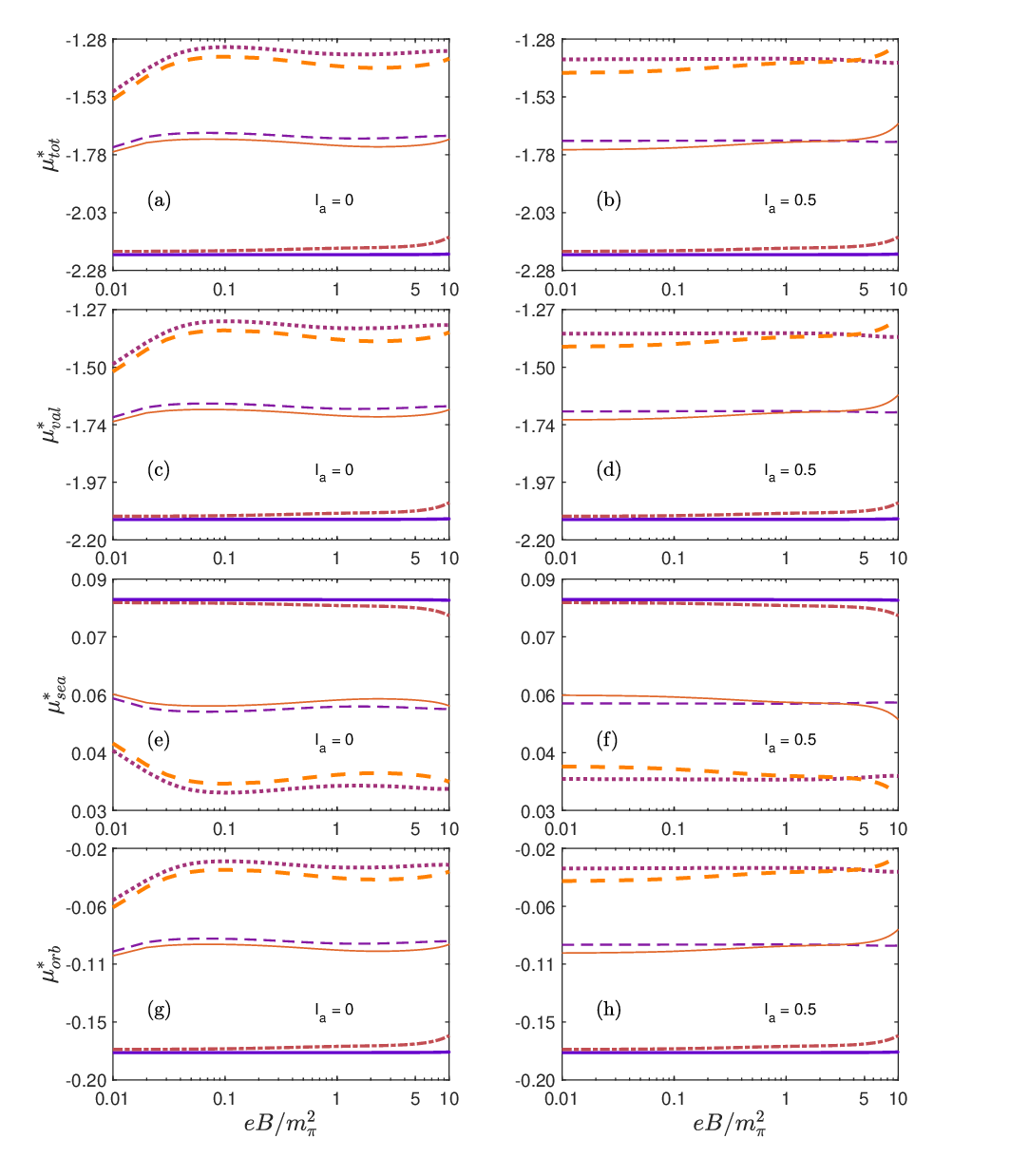}\\[-2ex]
	\includegraphics[width=0.55\textwidth]{Fig_legends.eps} \\[-3ex]
	\caption{The individual magnetic moment contributions from valence, sea and orbital moment of sea quarks to the total magnetic moment of \(\Xi^{*-}\) for \(I_a = 0\) (left hand panel) and \(I_a = 0.5\) (right hand panel) shown as a function of magnetic field from 0 to 10 \(eB/m_\pi^2\) and baryonic densities $\rho_B \,=\,0,\,\rho_0$ and $2\,\rho_0$.}
	\label{fig:XiM_magmom}
\end{figure}

\begin{figure}[H]
	\centering
	\includegraphics[width=1\textwidth]{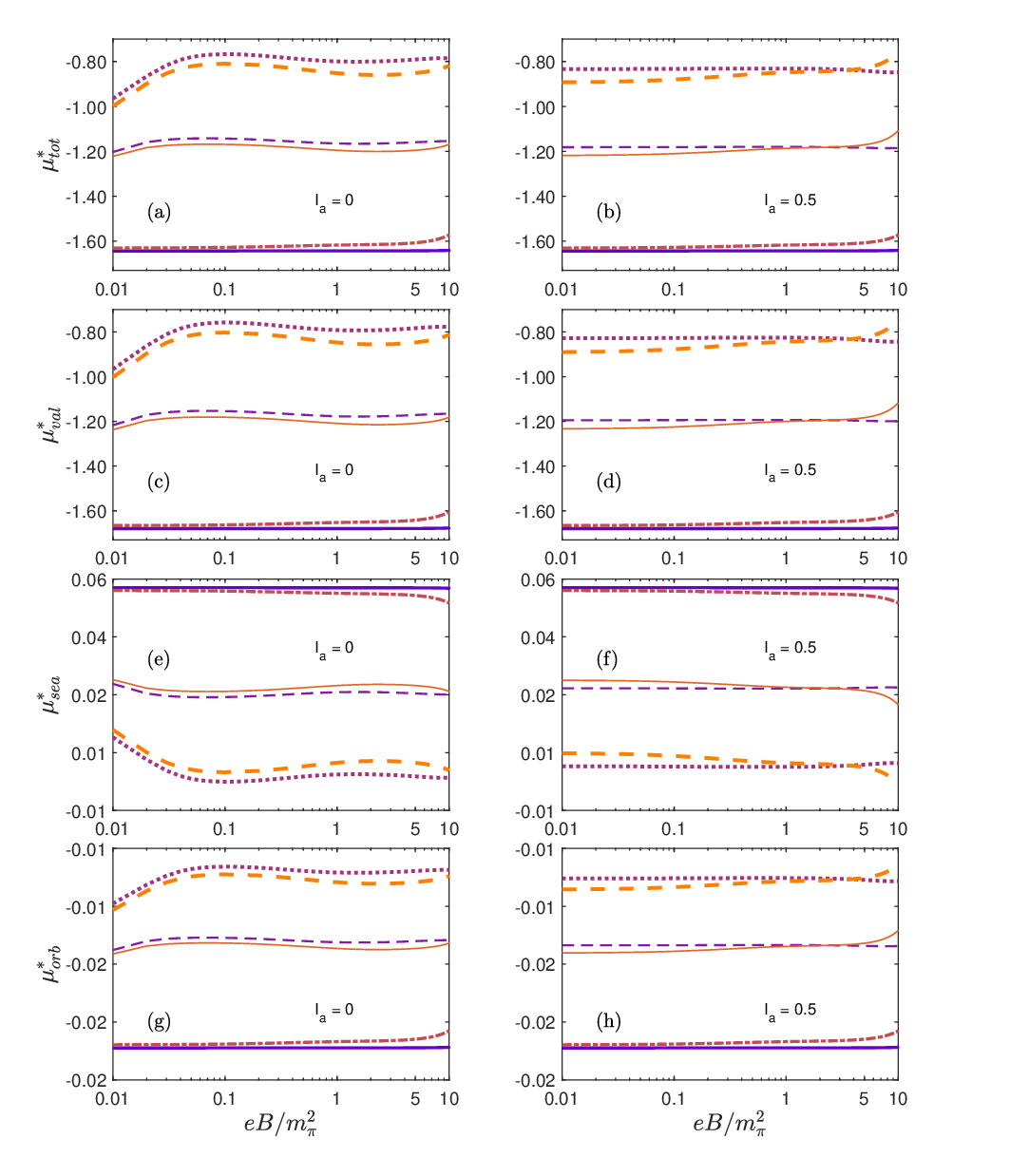}\\[-2ex]
	\includegraphics[width=0.55\textwidth]{Fig_legends.eps} \\[-3ex]
	\caption{The individual magnetic moment contributions from valence, sea and orbital moment of sea quarks to the total magnetic moment of \(\Omega^-\) for \(I_a = 0\) (left hand panel) and \(I_a = 0.5\) (right hand panel) shown as a function of magnetic field from 0 to 10 \(eB/m_\pi^2\) and baryonic densities $\rho_B \,=\,0,\,\rho_0$ and $2\,\rho_0$.}
	\label{fig:omega_magmom}
\end{figure}

	This section comprises the results of magnetic moments of decuplet baryons and their behaviour as affected by the landau quantization of fields in magnetized matter. Illustrations of the total effective magnetic moment ($\mu_{tot}^*$) as a function of magnetic field ($eB/m_\pi^2$), in units of $\mu_{N}$ (nuclear magneton), is shown at various densities $\rho_B \,=\,0,\, \rho_0$ and $ 2\,\rho_0$, isospin asymmetries $I_a=0$, $0.5$ and temperatures T = 100, 150 MeV. The total effective magnetic moments and the individual contribution consisting of valence, sea and orbital angular momentum of sea quarks for decuplet baryons are shown in Figs. \cref{fig:DeltaPP_magmom,fig:DeltaP_magmom,fig:Delta0_magmom,fig:DeltaM_magmom,fig:SigmaP_magmom,fig:Sigma0_magmom,fig:SigmaM_magmom,fig:Xi0_magmom,fig:XiM_magmom,fig:omega_magmom}. The changes in the masses of the baryons with respect to the external magnetic field are also reflected in the behaviour of magnetic moments of the baryons. 
	
	In the Figs. \cref{fig:DeltaPP_magmom,fig:DeltaP_magmom,fig:SigmaP_magmom} for temperature T = 100 MeV at density $\rho_B\,=\,0$ and $I_a\,=\,0$ [subplots (a), (c), (e) and (g)] the total effective magnetic moment ($\mu^*_{tot}$) for positively charged baryons $\Delta^{++}$, $\Delta^+$ and $\Sigma^{*+}$ are shown in subplots (a) for $I_a\,=\,0$. In the absence of external magnetic field with increase in baryonic density from zero to $\rho_B\,=\,\rho_0$ and further to $2\,\rho_0$ a consistent decrease in $\mu^*_{tot}$ was observed. As shown for the case of quark masses (Fig. \cref{fig:qm}) and baryon masses (Fig. \cref{fig:Xi_mass_log}) at $I_a\,=\,0$ the same is reflected here through the Eq. \eqref{eq:indivitual_quark_magnetic_moments}. For $\rho_B\,=\,0$ and $\rho_0$ increasing the magnetic field to $10\, m_\pi^2$ shows negligible change for the total baryonic magnetic moment while at $2\,\rho_0$ a significant decrease is observed for the positively charged baryons till $0.07\, m_\pi^2$ with no change observed thereafter with increase in magnetic field. The overall decrease in total magnetic moment for increase in magnetic field from zero to $10\,m_\pi^2$ at $2\,\rho_0$ for T = 100 MeV is observed to be $12\,\%$, $11\,\%$ and $6\,\%$ for $\Delta^{++},\,\Delta^+$ and $\Sigma^{*+}$ respectively. This indicates that heavier baryons show a smaller percentage decrease in their magnetic moments. The trend observed for temperature T = 150 MeV is found to be similar to that of T = 100 MeV with respect to magnetic field and no appreciable change was observed for magnetic moment with increment in temperature at a particular baryonic density. The total magnetic moment remains fairly similar when the isospin asymmetry factor is increased from zero to $I_a=0.5$ [subplots (b), (d), (f) and (h)] as $\mu^*_{tot}$ of $\Delta^{++}$, $\Delta^+$ and $\Sigma^{*+}$ decreases with increase in density from $\rho_B\,=\,0$ to $\rho_0$ and further to $2\,\rho_0$ at T = 100 MeV and zero magnetic field, similar to the case of $I_a\,=\,0$. The magnetic moments show almost no significant change with increasing magnetic field for densities $\rho_B\,=\,0$ and $\rho_0$ at T = 100 MeV. However, the dip in $\mu^*_{tot}$ that was observed for $I_a\,=\,0$ at $0.07\,m_\pi^2$ and density $\rho_B\,=\,2\,\rho_0$, disappears for increased $I_a\,=\,0.5$. At a fixed finite density  ($\rho_0$ and $2\,\rho_0$) and $I_a\,=\,0.5$ the trends for T = 100 MeV and 150 MeV are similar with respect to increasing magnetic field, showing a slight dependence on temperature. However, T = 150 MeV yields a higher value of magnetic moment compared to T = 100 MeV at magnetic fields below $2\,m_\pi^2$ followed by a decreasing magnetic moment till $10\,m_\pi^2$. 
	The calculation of total baryonic magnetic moment in the present work incorporates the valence, sea and orbital contributions. Therefore, the overall dependence of magnetic moment on the magnetized nuclear matter is dictated by the nature of these three distinct components. The valence quark magnetic moment, i.e., $\mu^*_{val}$ as shown in subplots (c) and (d) is the major contributing factor and shows similar nature to that of $\mu^*_{tot}$. For the positive baryons $\mu^*_{sea}$ [subplots (e) and (f)] shows lower negative values with increasing baryonic density at any particular temperature and magnetic field strength whereas a comparable but opposite nature is observed for $\mu^*_{orb}$ [subplots (g) and (h)] showing a decreasing positive value for increase in baryonic density.   
	
	The magnetic moment of neutral baryons, $\Delta^0,\,\Sigma^{*0}$ and $\Xi^{*0}$ are shown in Figs. \cref{fig:Delta0_magmom,fig:Sigma0_magmom,fig:Xi0_magmom} respectively. An increase in the total effective magnetic moment at T = 100 with increase in baryonic density was observed for the neutral baryons in symmetric nuclear matter. Increasing magnetic field to $10\,m_\pi^2$ yields no significant change for $\rho_B\,=\,0$ and $\rho_0$ densities at T = 100 MeV. Further increasing the density to $2\,\rho_0$ a significant increase in $\mu^*_{tot}$ was observed, recording an increase of $22\,\%,\, 11\,\%$ and $\,6\,\%$ for increase in magnetic field from zero to $0.07\,m_\pi^2$ for $\Delta^0,\, \Sigma^{*0}$ and $\Xi^{*0}$, respectively. The magnetic moment varies insignificantly with respect to increase in temperature from T = 100 MeV to 150 MeV as the two trends seem to be very similar. Increasing isospin asymmetry factor, $I_a\,=\,0.5$, at a fixed density and temperature the trends show no appreciable change with respect to increase in magnetic field. At density $\rho_B\,=\,2\,\rho_0$ for $I_a\,=\,0.5$ there is no increase as such observed for symmetric nuclear matter. At particular baryonic density with magnetic fields below $3\,m_\pi^2$, for neutrally charged baryons, higher temperature is observed to have a lower value of total magnetic moment, further increasing of magnetic field results in a reversal of the trend. For the case of $\Delta^0$ the major contribution is from $\mu^*_{sea}$ with $\mu^*_{val}$ contribution being null in both symmetric and asymmetric nuclear matter. The major contributions toward the total magnetic moment of baryons $\Sigma^{*0}$ and $\Xi^{*0}$ are from $\mu^*_{val}$ and $\mu^*_{sea}$ tend to show similar trends for both $I_a\,=\,0$ and 0.5. The total magnetic moment is found to have less positive value than that of $\mu^*_{val}$ as $\mu^*_{sea}$ tends to have a negative value at any particular magnetic field and density, thereby decreasing the $\mu^*_{tot}$ for $\Sigma^{*0}$ and $\Xi^{*0}$. The contribution from $\mu^*_{orb}$ has the lowest magnitude among all three individual contributing factors.
	
	The total effective magnetic moment of negatively charged baryons in Figs. \cref{fig:DeltaM_magmom,fig:SigmaM_magmom,fig:XiM_magmom,fig:omega_magmom} show an increase (decrease in magnitude) with increase in density from zero to $\rho_B\,=\,\rho_0$ at zero magnetic field and temperature T = 100 MeV in symmetric nuclear matter [subplot(a)]. At low densities ($\rho_B\,=\,0$ and $\rho_0$), the magnetic moment remains approximately constant as magnetic field is increased till $10\,m_\pi^2$. The negatively charged baryons  $\Delta^-$ (Fig. \cref{fig:DeltaM_magmom}) $,\, \Sigma^{*-}$ (Fig. \cref{fig:SigmaM_magmom}) $,\,\Xi^{*-}$ (Fig. \cref{fig:XiM_magmom}) and $\Omega^-$ (Fig. \cref{fig:omega_magmom}) for density of $2\,\rho_0$ and T = 100 MeV shows similar trends to that of positive and neutral baryons, showing a gradual increase of $14\,\%,\,11\,\%, \,13\,\%$ and $20\,\%$ for increasing magnetic field till $0.07 \, m_\pi^2$, thereafter the value remains almost constant till $10 \, m_\pi^2$.
	This gradual change observed for $2\,\rho_0$ density till $0.07m_\pi^2$ for $I_a\,=\,0$, vanishes with increase in isospin asymmetry to $I_a\,=\,0.5$. The valence contribution, $\mu^*_{val}$, is the major contributing factor to the total magnetic moment showing increase in value with increase in density from $\rho_B\,=\,0$ to $\rho_0$ and $2\,\rho_0$ at zero magnetic field. In contrast to the other contributing factors of magnetic moment, $\mu^*_{sea}$ shows opposite trends of decreasing value (becomes less positive) with increasing density at T = 100 MeV at zero magnetic field. For the case of negatively charged baryons the $\mu^*_{orb}$ contribution outweighs the values of the $\mu^*_{sea}$ at particular magnetic field and density. As one can see from \cref{tab:t100_0mf_eta0_magmoment,tab:t100_0mf_eta0.5_magmoment}, for the heavier baryons $\Xi^{*-}$ and $\Omega^-$ the values of $\mu^*_{sea}$ and $\mu^*_{orb}$ have approximately equal values in magnitude but opposite in sign and hence almost neutralize each other. As a result, the magnetic moment of these baryons is majorly influenced by the contributions from the $\mu^*_{val}$.

\vspace{-1em}

\begin{table}[H]
	\begingroup
	\footnotesize
	\caption{Total magnetic moments and the respective contributions from valence quarks, sea quarks, and the orbital magnetic moment of sea quarks at different baryonic densities at T = 100 MeV, $I_a$ = 0 and $eB \,=\,0\, m_\pi^2$.}
	\centering
	\renewcommand{\arraystretch}{0.8}
	\begin{tabular}{|c|c|c|c|c|c|c|c|c|c|c|c|c|} 
		\hline
		\multirow{2}{*}{\centering$\begin{array}{c}
				\mu^*(\mu_N) \\
				T = 100\,\mathrm{MeV}
			\end{array}$} &
		\multicolumn{4}{c|}{$\rho_B = 0$} &
		\multicolumn{4}{c|}{$\rho_B = \rho_0$} &
		\multicolumn{4}{c|}{$\rho_B = 2\,\rho_0$} \\
		\cline{2-5}\cline{6-9}\cline{10-13}
		& $\mu^*_{\text{total}}$ & $\mu^*_{\text{val}}$ & $\mu^*_{\text{sea}}$ & $\mu^*_{\text{orbital}}$ 
		& $\mu^*_{\text{total}}$ & $\mu^*_{\text{val}}$ & $\mu^*_{\text{sea}}$ & $\mu^*_{\text{orbital}}$ 
		& $\mu^*_{\text{total}}$ & $\mu^*_{\text{val}}$ & $\mu^*_{\text{sea}}$ & $\mu^*_{\text{orbital}}$\\
		\hline
		$\mu_{\Delta^{++}}^*$ & 5.482 & 5.982 & $-1.048$ & 0.547 & 4.461 & 5.008 & $-0.884$ & 0.337 & 3.871 & 4.419 & $-0.783$ & 0.235 \\ \hline
		
		$\mu_{\Delta^{+}}^*$ & 2.538 & 2.991 & $-0.657$ & 0.205 & $2.084$ & $2.503$ & $-0.556$ & $0.137$ & $1.819$ & $3.367$ & $2.206$ & $-0.494$ \\ \hline
		
		$\mu_{\Delta^{0}}^*$& $-0.405$ & 0 & $-0.267$ & $-0.138$ & $-0.292$ & 0 & $-0.229$ & $-0.063$ & $-0.228$ & 0 & $-0.205$ & $-0.022$  \\ \hline
		
		$\mu_{\Delta^{-}}^*$& $-3.348$ & $-2.991$ & 0.124 & $-0.481$ & $-2.667$ & $-2.502$ & 0.098 & $-0.263$ & -2.268 & -2.199 & 0.082 & -0.151  \\ \hline
		
		$\mu_{\Sigma^{*+}}^*$& $3.109$ & $3.431$ & $-0.679$ & $0.358$ & $2.755$ & $3.149$ & $-0.614$ & $0.219$ & 2.567 & 2.994 & -0.577 & 0.153   \\ \hline
		
		$\mu_{\Sigma^{*0}}^*$& $0.164$ & $0.478$ & $-0.289$ & $0.015$ & $0.254$& $0.503$ & $-0.268$ & $0.020$ & 0.300 & 0.532 & -0.256 & 0.024  \\ \hline 
		
		$\mu_{\Sigma^{*-}}^*$& $-2.780$ & $-2.554$ & $0.102$ & $-0.328$ & $-2.245$ & $-2.143$ & $0.078$ & $-0.180$ & $-1.961$ & $-1.923$ & 0.064 & $-0.104$  \\ \hline
		
		$\mu_{\Xi^{*0}}^*$   & $0.733$ & $0.876$ & $-0.311$ & 0.168 & $0.842$ & $1.044$ & $-0.305$ & $0.1203$  & 0.907 & 1.139 & $-0.303$ & 0.071  \\ \hline
		
		$\mu_{\Xi^{*-}}^*$   & $-2.212$ & $-2.117$ & 0.079 & $-0.174$ & $-1.747$ & $-1.705$ & 0.054 & $-0.097$ & $-1.507$ & $-1.490$ & 0.041 & $-0.058$   \\ \hline 
		
		$\mu_{\Omega^{-}}^*$ & $-1.643$ & $-1.679$ & 0.057 & $-0.020$ & $-1.202$ & $-1.217$ & 0.028 & $-0.014$ & $-0.966$ & $-0.967$ & 0.012 & -0.011   \\  
		\hline
	\end{tabular}
	
	\label{tab:t100_0mf_eta0_magmoment}
	\endgroup
\end{table}
\vspace{-1.5em}
\small

\begin{table}[H]
	\begingroup
	\footnotesize
	\caption{Total magnetic moments and the respective contributions from valence quarks, sea quarks, and the orbital magnetic moment of sea quarks at different baryonic densities at T = 100 MeV, $I_a$ = 0 and $eB \,=\,10\, m_\pi^2$.}
	\centering
	\renewcommand{\arraystretch}{0.8}
	\begin{tabular}{|c|c|c|c|c|c|c|c|c|c|c|c|c|} 
		\hline
		\multirow{2}{*}{\centering$\begin{array}{c}
				\mu^*(\mu_N) \\
				T = 100\,\mathrm{MeV}
			\end{array}$} &
		\multicolumn{4}{c|}{$\rho_B = 0$} &
		\multicolumn{4}{c|}{$\rho_B = \rho_0$} &
		\multicolumn{4}{c|}{$\rho_B = 2\,\rho_0$} \\
		\cline{2-5}\cline{6-9}\cline{10-13}
		& $\mu^*_{\text{total}}$ & $\mu^*_{\text{val}}$ & $\mu^*_{\text{sea}}$ & $\mu^*_{\text{orbital}}$ 
		& $\mu^*_{\text{total}}$ & $\mu^*_{\text{val}}$ & $\mu^*_{\text{sea}}$ & $\mu^*_{\text{orbital}}$ 
		& $\mu^*_{\text{total}}$ & $\mu^*_{\text{val}}$ & $\mu^*_{\text{sea}}$ & $\mu^*_{\text{orbital}}$\\
		\hline
		$\mu_{\Delta^{++}}^{\!*}$ & 5.476 & 5.977 & $-1.047$ & 0.546 & 4.342 & 4.890 & $-0.864$ & 0.315 & 3.407 & 3.944 & -0.701 & 0.164 \\ \hline
		
		$\mu_{\Delta^{+}}^{\!*}$  & 2.536 & 2.988 & $-0.657$ & 0.204 & 2.031 & 2.446 & -0.544 & 0.130 & 1.614 & 1.973 & -0.444 & 0.085 \\ \hline
		
		$\mu_{\Delta^{0}}^{\!*}$  & $-0.404$ & 0 & $-0.266$ & $-0.138$ & $-0.280$ & 0 & $-0.225$ & $-0.055$ & -0.181 & 0 & $-0.186$ & 0.005 \\ \hline
		
		$\mu_{\Delta^{-}}^{\!*}$  & $-3.344$ & $-2.988$ & 0.124 & $-0.480$ & $-2.592$ & $-2.446$ & 0.095 & $-0.240$ & -1.979 & -1.976 & 0.072 & -0.075 \\ \hline
		
		$\mu_{\Sigma^{*+}}^{\!*}$  & 3.106 & 3.429 & -0.679 & 0.357 & 2.715 & 3.115 & -0.606 & 0.206 & 2.419 & 2.860 & -0.548 & 0.107 \\ \hline
		
		$\mu_{\Sigma^{*0}}^{\!*}$  &  0.165 & 0.438 & -0.289 & 0.015  & 0.264 & 0.509 & -0.266 & 0.021 & 0.330 & 0.549 & -0.246 & 0.027  \\ \hline
		
		$\mu_{\Sigma^{*-}}^{\!*}$  & $-2.777$ & $-2.552$ & 0.102 & $-0.327$  & $-2.188$ & $-2.099$ & 0.075 & $-0.165$ & $-1.760$ & $-1.763$ & 0.056 & $-0.053$  \\ \hline
		
		$\mu_{\Xi^{*0}}^{\!*}$     & 0.734 & 0.877 & -0.311 & 0.168  & 0.855 & 1.063 & -0.304 & 0.096 & 0.954 & 1.206 & -0.301 & 0.049  \\ \hline
		
		$\mu_{\Xi^{*-}}^{\!*}$     & -2.209 & -2.115 & 0.080 & -0.174  & -1.698 & -1.660 & 0.051 & -0.089 & -1.332 & -1.332 & 0.031 & -0.031 \\ \hline
		
		$\mu_{\Omega^{-}}^{\!*}$  & -1.641 & -1.677 & 0.057 & -0.021  & -1.154 & -1.166 & 0.025 & -0.013 & -0.785 & -0.776 & -0.000 & -0.008  \\ 
		\hline
	\end{tabular}
	
	\label{tab:t100_10mf_eta0_magmoment}
	\endgroup
\end{table}

\begin{table}[H]
	\begingroup
	\footnotesize
	\caption{Total magnetic moments and the respective contributions from valence quarks, sea quarks, and the orbital magnetic moment of sea quarks at different baryonic densities at T = 100 MeV, $I_a$ = 0.5 and $eB \,=\,0\, m_\pi^2$.}
	\centering
	\renewcommand{\arraystretch}{0.8}
	\begin{tabular}{|c|c|c|c|c|c|c|c|c|c|c|c|c|} 
		\hline
		\multirow{2}{*}{\centering$\begin{array}{c}
				\mu^*(\mu_N) \\
				T = 100\,\mathrm{MeV}
			\end{array}$} &
		\multicolumn{4}{c|}{$\rho_B = 0$} &
		\multicolumn{4}{c|}{$\rho_B = \rho_0$} &
		\multicolumn{4}{c|}{$\rho_B = 2\,\rho_0$} \\
		\cline{2-5}\cline{6-9}\cline{10-13}
		& $\mu^*_{\text{total}}$ & $\mu^*_{\text{val}}$ & $\mu^*_{\text{sea}}$ & $\mu^*_{\text{orbital}}$ 
		& $\mu^*_{\text{total}}$ & $\mu^*_{\text{val}}$ & $\mu^*_{\text{sea}}$ & $\mu^*_{\text{orbital}}$ 
		& $\mu^*_{\text{total}}$ & $\mu^*_{\text{val}}$ & $\mu^*_{\text{sea}}$ & $\mu^*_{\text{orbital}}$\\
		\hline
		$\mu_{\Delta^{++}}^*$ & 5.482 & 5.982 & $-1.048$ & 0.547 & 4.482 & 4.956 & $-0.875$ & 0.327 & 3.533 & 4.074 & $-0.724$ & 0.183 \\ \hline
		
		$\mu_{\Delta^{+}}^*$ & 2.538 & 2.991 & $-0.657$ & 0.205 & $2.056$ & $2.469$ & $-0.549$ & $0.136$ & $1.660$ & $2.019$ & $-0.454$ & $0.095$ \\ \hline
		
		$\mu_{\Delta^{0}}^*$& $-0.405$ & 0 & $-0.267$ & $-0.138$ & $-0.282$ & 0 & $-0.226$ & $-0.056$ & $-0.182$ & 0 & $-0.188$ & $0.006$  \\ \hline
		
		$\mu_{\Delta^{-}}^*$& $-3.348$ & $-2.991$ & 0.124 & $-0.481$ & $-2.605$ & $-2.453$ & 0.095 & $-0.248$ & $-1.993$ & $-1.984$ & 0.073 & $-0.082$  \\ \hline
		
		$\mu_{\Sigma^{*+}}^*$& 3.109 & 3.430 & $-0.680$ & 0.358 & 2.735 & 3.131 & $-0.610$ & 0.214 & 2.454 & 2.891 & $-0.555$ & 0.119   \\ \hline
		
		$\mu_{\Sigma^{*0}}^*$& 0.164 & 0.438 & -0.289 & 0.015  & 0.259 & 0.503 & -0.266 & 0.022 & 0.324 & 0.540 & -0.247 & 0.031  \\ \hline 
		
		$\mu_{\Sigma^{*-}}^*$& 2.780 & $-2.555$ & 0.102 & $-0.328$ & $-2.205$ & $-2.112$ & 0.076 & $-0.170$ & $-1.783$ & $-1.783$ & 0.058 & $-0.057$  \\ \hline
		
		$\mu_{\Xi^{*0}}^*$   & 0.733 & 0.876 & $-0.311$ & 0.169 & 0.846 & 1.050 & $-0.304$ & 0.100 & 0.938 & 1.184 & $-0.301$ & 0.055  \\ \hline
		
		$\mu_{\Xi^{*-}}^*$   & $-2.212$ & $-2.117$ & 0.080 & $-0.174$ & $-1.720$ & $-1.681$ & 0.053 & $-0.092$ & $-1.368$ & $-1.367$ & 0.033 & $-0.033$    \\ \hline  
		
		$\mu_{\Omega^{-}}^*$ & $-1.643$ & $-1.680$ & 0.057 & $-0.021$ & $-1.181$ & $-1.195$ & 0.027 & $-0.014$ & $-0.834$ & $-0.828$ & 0.003 & $-0.009$   \\  
		\hline
	\end{tabular}
	
	\label{tab:t100_0mf_eta0.5_magmoment}
	\endgroup
\end{table}
\vspace{-1.5em}
\small
\begin{table}[H]
	\begingroup
	\footnotesize
	\caption{Total magnetic moments and the respective contributions from valence quarks, sea quarks, and the orbital magnetic moment of sea quarks at different baryonic densities at T = 100 MeV, $I_a$ = 0.5 and $eB \,=\,10\, m_\pi^2$.}
	\centering
	\renewcommand{\arraystretch}{0.8}
	\begin{tabular}{|c|c|c|c|c|c|c|c|c|c|c|c|c|} 
		\hline
		\multirow{2}{*}{\centering$\begin{array}{c}
				\mu^*(\mu_N) \\
				T = 100\,\mathrm{MeV}
			\end{array}$} &
		\multicolumn{4}{c|}{$\rho_B = 0$} &
		\multicolumn{4}{c|}{$\rho_B = \rho_0$} &
		\multicolumn{4}{c|}{$\rho_B = 2\,\rho_0$} \\
		\cline{2-5}\cline{6-9}\cline{10-13}
		& $\mu^*_{\text{total}}$ & $\mu^*_{\text{val}}$ & $\mu^*_{\text{sea}}$ & $\mu^*_{\text{orbital}}$ 
		& $\mu^*_{\text{total}}$ & $\mu^*_{\text{val}}$ & $\mu^*_{\text{sea}}$ & $\mu^*_{\text{orbital}}$ 
		& $\mu^*_{\text{total}}$ & $\mu^*_{\text{val}}$ & $\mu^*_{\text{sea}}$ & $\mu^*_{\text{orbital}}$\\
		\hline
		$\mu_{\Delta^{++}}^*$ & 5.476 & 5.977 & $-1.047$ & 0.546 & 4.416 & 4.963 & $-0.876$ & 0.329 & 3.569 & 4.111 & $-0.730$ & 0.188 \\ \hline 
		
		$\mu_{\Delta^{+}}^*$ & 2.538 & 2.991 & $-0.657$ & 0.205 & $2.056$ & $2.469$ & $-0.549$ & $0.136$ & $1.660$ & $2.019$ & $-0.454$ & $0.095$ \\ \hline
		
		$\mu_{\Delta^{0}}^*$& $-0.405$ & 0 & $-0.267$ & $-0.138$ & $-0.282$ & 0 & $-0.226$ & $-0.056$ & $-0.182$ & 0 & $-0.188$ & $0.006$  \\ \hline
		
		$\mu_{\Delta^{-}}^*$& $-3.348$ & $-2.991$ & 0.124 & $-0.481$ & $-2.605$ & $-2.453$ & 0.095 & $-0.248$ & $-1.993$ & $-1.984$ & 0.073 & $-0.082$  \\ \hline
		
		$\mu_{\Sigma^{*+}}^*$& 3.109 & 3.430 & $-0.680$ & 0.358 & 2.735 & 3.131 & $-0.610$ & 0.214 & 2.454 & 2.891 & $-0.555$ & 0.119   \\ \hline
		
		$\mu_{\Sigma^{*0}}^*$& 0.164 & 0.438 & -0.289 & 0.015  & 0.259 & 0.503 & -0.266 & 0.022 & 0.324 & 0.540 & -0.247 & 0.031  \\ \hline 
		
		$\mu_{\Sigma^{*-}}^*$& 2.780 & $-2.555$ & 0.102 & $-0.328$ & $-2.205$ & $-2.112$ & 0.076 & $-0.170$ & $-1.783$ & $-1.783$ & 0.058 & $-0.057$  \\ \hline
		
		$\mu_{\Xi^{*0}}^*$   & 0.733 & 0.876 & $-0.311$ & 0.169 & 0.846 & 1.050 & $-0.304$ & 0.100 & 0.938 & 1.184 & $-0.301$ & 0.055  \\ \hline
		
		$\mu_{\Xi^{*-}}^*$   & $-2.212$ & $-2.117$ & 0.080 & $-0.174$ & $-1.720$ & $-1.681$ & 0.053 & $-0.092$ & $-1.368$ & $-1.367$ & 0.033 & $-0.033$    \\ \hline  
		
		$\mu_{\Omega^{-}}^*$ & $-1.643$ & $-1.680$ & 0.057 & $-0.021$ & $-1.181$ & $-1.195$ & 0.027 & $-0.014$ & $-0.834$ & $-0.828$ & 0.003 & $-0.009$   \\  
		\hline
	\end{tabular}
	
	\label{tab:t100_10mf_eta0.5_magmoment}
	\endgroup
\end{table}

	\section{Summary} \label{summary}
	The impact of magnetic fields on the masses of baryons has been studied through the chiral SU(3) quark mean field model by incorporating the landau quantization. We obtain the in-medium modified masses of decuplet baryons in isospin asymmetric magnetized nuclear matter. The results show that baryonic density as well as temperatures and isospin asymmetry factor tend to impact the quark masses, baryon masses and the magnetic moment of decuplet baryons as well.
	
	A significant change in the quark masses and thus the magnetic moment of each baryon is oberserved around $0.07\,m_\pi^2$ magnetic field which is more prominent for symmetric nuclear matter $I_a\,=\,0$ at finite baryonic densities. With further increase in the magnetic field, changes observed for the magnetic moment of the baryons are insignificant till $10\,m_\pi^2$ for both symmetric and asymmetric nuclear matter.	
	For the case of asymmetric nuclear matter $I_a\,=\,0.5$, the variation caused by weak magnetic fields is suppresed and a change is observed with increasing baryonic density. The temperatures tend to coalesce together for both the temperatures at any particular baryonic density till $3\,m_\pi^2$. Our work is expected to be useful for future experimental facilities where hot dense nuclear matter under intense magnetic field is produced and studied. The masses and magnetic moments are expected to aid in the exploration of internal structure of baryons further. 
	
	
	\bibliographystyle{apsrev4-1}

\end{document}